# An extended Rice model for intergranular fracture

Kai Zhao[1,2*], Yu Ding[3], Haiyang Yu[4], Jianying He[3], Zhiliang Zhang[3*]

*1 School of Mechanical Engineering, Jiangnan University, Wuxi 214122, China*

*2 Jiangsu Key Laboratory of Advanced Food Manufacturing Equipment and Technology, Wuxi 214122, China*

*3 Department of Structural Engineering, Norwegian University of Science and Technology (NTNU), Trondheim 7491, Norway*

*4 Division of Applied Mechanics, Department of Materials Science and Engineering, Uppsala University, Uppsala 75121, Sweden*

## Abstract

The plastic events occurring during the process of intergranular fracture in metals is still not well understood due to the complexity of grain boundary (GB) structures and their interactions with crack-tip dislocation plasticity. By considering the local GB structural transformation after dislocation emission from a GB in the Peierls-type Rice-Beltz model, herein we established a semi-analytical transition-state-theory-based framework to predict the most probable Mode-I stress intensity factor (SIF) for dislocation emission from a cracked GB. Using large-scale molecular dynamics (MD) simulations, we studied the fracture behaviors of bi-crystalline Fe samples with 12 different symmetric tilt GBs inside. The MD results demonstrate that the presence of GB could significantly change the SIF required for the activation of plastic events, confirming the theoretical predictions that attributes this to the energy change caused by the transformation of GB structure. Both the atomistic simulation and the theoretical model consistently indicate that, the critical dynamic SIF ($K_I^c(t)$) at which the dynamic SIF $K_I(t)$ deviates from the linearity with respect to the strain $\varepsilon$, increases with the increasing loading rate. However, the classical Rice model underestimates the $K_I^c(t)$ due to its failure to consider the effects of localized fields. The present theoretical model provides a mechanism-based framework for the application of grain boundary engineering in the design and fabrication of nano-grained metals.

[*] Corresponding authors: kai.zhao@jiangnan.edu.cn (K. Z.), zhiliang.zhang@ntnu.no (Z. Z.)

## 1. Introduction

With the characteristic length scale of microstructures on the order of a few (typically 1-10) nanometers [1], nanostructured materials (NSMs) have been receiving considerable attentions in past decades due to their extraordinary performance in mechanical, electrical and optical applications [2]. Although the materials strength can be drastically enhanced by reducing the grain size (*i.e.* the Hall-Petch law [3, 4]), a large fraction of grain boundaries (GBs) serve as the natural sites of cleavage fracture [5] and promote the nucleation of dislocations. Whether the fracture occurs by ductile rupture or by brittle cleavage is determined by the competition between the bond-breaking at the crack-tip and the plastic deformation in the vicinity of the crack [6-9]. Therefore, to better understand and design NSMs, it is necessary to illustrate the effect of GBs on the variation of crack patterns.

It is widely accepted that whether the material is intrinsically ductile or brittle in terms of the atomic structure at the tip of a sharp crack, is determined by the competition of two cracking modes: the material is ductile if the crack under external loading was blunted by dislocation nucleation rather than cleaved by crack propagation (*vice versa*) [10]. Theoretical analysis of this competition has been well established [11-14], e.g. the continuum models developed by Kelly *et al*. [15] and later Rice and Tomson [16]. By incorporating the Peierls framework into the dislocation nucleation description, Rice [12] showed that under the Mode II (in-plane shear) loading the dislocation emission is controlled by an energy criterion involving the unstable stacking fault (USF) energy $\gamma_{usf}$. Subsequent studies have advanced the Rice framework and the Rice-Thomson model by accounting for the elastic anisotropy [14], the effect of crack blunting [17, 18], three-dimensional dislocation nuclei [13], successive nucleation events [19], and surface steps [20] formed at the crack-tip [10].

However, with the presence of GBs, the crack-tip behavior might be influenced by either the dislocation emission from GBs or deformation correlated to GB structures [21-25]. Möller and Bitzek [8] studied the atomic-scale fracture behavior of large-angle tilt GBs in bcc W bicrystals, and found that the fracture toughness critically depends on the propagation direction and the position of the crack-tip within the structural units of the GB, *i.e.* the effect of bond trapping. Besides, the fracture toughness of GBs can be significantly larger than that of single crystals, and the maximum GB fracture toughness is not necessarily correlated with the GB energy. Cheng *et al*. [26] investigated the intrinsic brittleness and ductility of intergranular fracture along the [110] symmetric tilt grain boundaries (STGBs) in Cu bicrystals, and found that the directional anisotropy predicted by the Rice model [12] is validated for coherent $\Sigma3(1\bar{1}1)$ and $\Sigma11(1\bar{1}3)$ GBs, but not observed for incoherent GBs, such as $\Sigma9(2\bar{2}1)$, $\Sigma9(1\bar{1}4)$ and $\Sigma11(3\bar{3}2)$. They attributed this discrepancy to the dislocation emission at a distance ahead of the crack-tip along an incoherent GB, which deviates from the hypothesis that dislocations are directly emitted from the crack-tip in the Rice model [12]. Subsequently, Shimokawa and Tsuboi [5] rationalized this dislocation emission from a site on the GB in an improved Rice-Thomson model [16]. Their results for Al bicrystals showed that the intergranular fracture toughness is affected by both the energy and structure of GBs. It is thus necessary to clarify the competition between the crack-tip and GBs as the emission source of dislocations during the fracture process.

Since the crack-tip plasticity is usually not negligible, the deformation patterns could significantly affect the crack propagation. Previous studies [27, 28] have demonstrated that the bcc lattice near the crack-tip could be deformed by fcc transition, twinning and dislocation nucleation depending on the crack geometry and loading orientation etc. While the bcc-fcc transition is usually metastable [29, 30], the competition between dislocation slip and twinning affect the crack-tip plasticity significantly [31-33]. Remington *et al*. [34] performed nanoindentation tests of bcc Ta, and found that the plastic deformation proceeds by the formation of nanotwins, which rapidly evolve into shear loops. Yamakov *et al*. [35] studied the dislocation nucleation processes near the crack-tip in fcc Al, and found that the partial dislocation nucleation could evolve into both full dislocation emission and twinning, depending on the orientation, temperature, and magnitude of the applied load. However, for the intergranular fracture process, it is not clear how the GBs affect the competition between dislocation emission and twinning in the vicinity of the crack-tip, which would subsequently affect the crack propagation. In other words, the contribution of the localized stress field induced by the plastic activities should be involved to compute the total energy for dislocation emission under the Rice framework [13].

While the localized plasticity at the cracked GBs is known to play a role in the fracture mode transition, correct treatments and proper considerations of it are crucial for understanding the duality of GBs in the design of nano-structured materials with both high toughness and strength. Currently, the available approaches based on the classical Rice model are unable to consider the localized events, thus their predictability should be scrutinized.

In this work, we present a novel framework to theoretically predict the critical Mode-I stress intensity factor (SIF, $K_I$) required for dislocation nucleation by extending the classical Rice-Beltz model [13] and considering the deformation of GB structural units near the crack-tip. In order to verify this new model, we also studied the interaction between the GBs and dislocations during the intergranular fracture process of a number of $[1\bar{1}0]$ tilt GBs in Fe bicrystals using large-scale atomistic simulations. The results show that the existence of GBs could significantly reduce or enhance the critical SIF required for the crack propagation due to the transformation of the GB structural unit. While the classical model fails to predict the enhancement of $K_I$ induced by the localized fields, the most probable SIF ($K_I^p$) derived by the present theory agree well with the large-scale MD simulation results.

## 2. Numerical protocols

In this section, the construction method of bi-crystalline specimens and MD simulation settings are described in details. By setting the $z$-axis along the $[1\bar{1}0]$ lattice direction as the tilt axis, 12 types of STGB with the tilt angle ranging from 20.05° to 148.4° were created. Then, a through-thickness crack was created to study the intergranular fracture behaviors under uniaxial tension.

### 2.1 Construction of bi-crystalline samples

Without loss of generality, following the approach proposed in Ref.s[36-39], the $\langle 110 \rangle$ STGBs covering the entire range of tilt angles are constructed by joining two crystal grains together and deleting overlapped atoms at the GB plane. The length of the computational cell along the GB normal should be large enough to eliminate any interaction between neighboring GBs. However, due to the variation of the relative translation with respect to each other grain and atomic densities at the interface [40], there exist multiple GB structures and energies for a single set of macroscopic degrees of freedom (DOF). By rigidly translating the two crystals relative to each other and restricting atom movement to the direction of the GB normal, the GB structure is optimized in terms of microscopic DOF through energy minimization. Then, the GB structure is relaxed without restriction of atomic motion, meanwhile allowing the variation of computational cell size in the $x$- and $z$-axis directions to obtain a stress-free sample (see **Fig. 1**a). Widely used in previous studies [40, 41], the conjugate gradient algorithm is chosen to obtain the minimum energy configuration. The lattice configurations of 12 different STGBs are shown in **Table 1**. It is noted that the Σ3(109.5) and Σ11(50.50) GBs are coherent twin boundaries, while the Σ3(70.00) GB is the incoherent twin boundary.

A through-thickness crack along the GB is then created by deleting atoms in the center of the $x$-$y$ plane, as shown in **Fig. 1**b. Considering the existence of the crack bluntness [42], the crack-tip with a given radius was created. According to Ref.[43], the Mode-I SIF $K_I$ can be correlated with the applied uniaxial tensile stress $\sigma$ as,

$$K_I = \sigma\sqrt{\pi a}F(a/b), \tag{1}$$

where, the empirical formula of $F(a/b)$ can be adopted as,

$$F(a/b) = \{1 - 0.025(a/b)^2 + 0.06(a/b)^4\}\sqrt{\sec\frac{\pi a}{2b}}, \tag{2}$$

Alternatively, one can also evaluate the $K_I$ via the exact solution for a periodic array of cracks in an infinite plate as [43],

$$K_I = \sigma\sqrt{W\tan\frac{\pi a}{W}}, \tag{3}$$

where, $W = 2b$ is the periodicity of the crack array. The difference between the results given by Eq.(1) and Eq.(3) is not significant, and can be ignored.

## 2.2 Simulation details

All atomistic simulations in present study are conducted by the open source MD code Large-scale Atomic/Molecular Massively Parallel Simulator (LAMMPS) [44]. The interatomic interaction between Fe atoms is described by using the Finnis-Sinclair type modification [45] of the embedded-atom-method (EAM) potential originally developed by Ramasubramaniam *et al*. [46] and modified by Song and Curtin [45]. The time step for the velocity-Verlet integration is set as 1 fs. After building the bicrystalline specimens via the procedures demonstrated in **Section 2.1**, the simulation system is created by deleting the atoms of the central region shown in **Fig. 1**a, then equilibrated at 300 K in the NPT ensemble with the *x*- and *z*-axis directions set as periodic, but the *y*-axis direction nonperiodic. Uniaxial tensile loading is then applied in the NVE ensemble along the *y*-axis direction. To output the profile of stress, temperature, density and particle velocity, the specimen is sliced into a series of bins along the *x*-axis direction. The dislocation extraction algorithm [47] is applied to identify dislocations generated during the loading process. The adaptive common neighbor analysis (a-CNA) pattern [48, 49] and centro-symmetry parameters [50] are calculated for post-processing. The defective structures are visualized in OVITO code [51].

**Table 1**. Lattice orientation of the 12 different STGBs built in present study.

| specimen index | GB index | grain #1 | | | grain #2 | | |
|---|---|---|---|---|---|---|---|
| | | *x*-axis | *y*-axis | *z*-axis | *x*-axis | *y*-axis | *z*-axis |
| S#1 | Σ3(70.00) | $[11\bar{1}]$ | [112] | $[1\bar{1}0]$ | [111] | $[\bar{1}\bar{1}2]$ | $[1\bar{1}0]$ |
| S#2 | Σ3(109.5) | $[11\bar{2}]$ | [111] | $[1\bar{1}0]$ | [112] | $[\bar{1}\bar{1}1]$ | $[1\bar{1}0]$ |
| S#3 | Σ9(38.90) | $[22\bar{1}]$ | [114] | $[1\bar{1}0]$ | [221] | $[\bar{1}\bar{1}4]$ | $[1\bar{1}0]$ |
| S#4 | Σ9(141.6) | [114] | $[\bar{2}\bar{2}1]$ | $[1\bar{1}0]$ | $[11\bar{4}]$ | [221] | $[1\bar{1}0]$ |
| S#5 | Σ11(50.50) | $[33\bar{2}]$ | [113] | $[1\bar{1}0]$ | [332] | $[\bar{1}\bar{1}3]$ | $[1\bar{1}0]$ |
| S#6 | Σ11(129.5) | [113] | $[\bar{3}\bar{3}2]$ | $[1\bar{1}0]$ | $[11\bar{3}]$ | [332] | $[1\bar{1}0]$ |
| S#7 | Σ17(86.60) | $[33\bar{4}]$ | [223] | $[1\bar{1}0]$ | [334] | $[\bar{2}\bar{2}3]$ | $[1\bar{1}0]$ |
| S#8 | Σ17(93.40) | $[22\bar{3}]$ | [334] | $[1\bar{1}0]$ | [223] | $[\bar{3}\bar{3}4]$ | $[1\bar{1}0]$ |
| S#9 | Σ19(26.50) | $[33\bar{1}]$ | [116] | $[1\bar{1}0]$ | [331] | $[\bar{1}\bar{1}6]$ | $[1\bar{1}0]$ |
| S#10 | Σ27(148.4) | $[11\bar{5}]$ | [552] | $[1\bar{1}0]$ | [115] | $[\bar{5}\bar{5}2]$ | $[1\bar{1}0]$ |
| S#11 | Σ33(20.05) | $[44\bar{1}]$ | [118] | $[1\bar{1}0]$ | [441] | $[\bar{1}\bar{1}8]$ | $[1\bar{1}0]$ |
| S#12 | Σ43(80.60) | $[55\bar{6}]$ | [335] | $[1\bar{1}0]$ | [556] | $[\bar{3}\bar{3}5]$ | $[1\bar{1}0]$ |

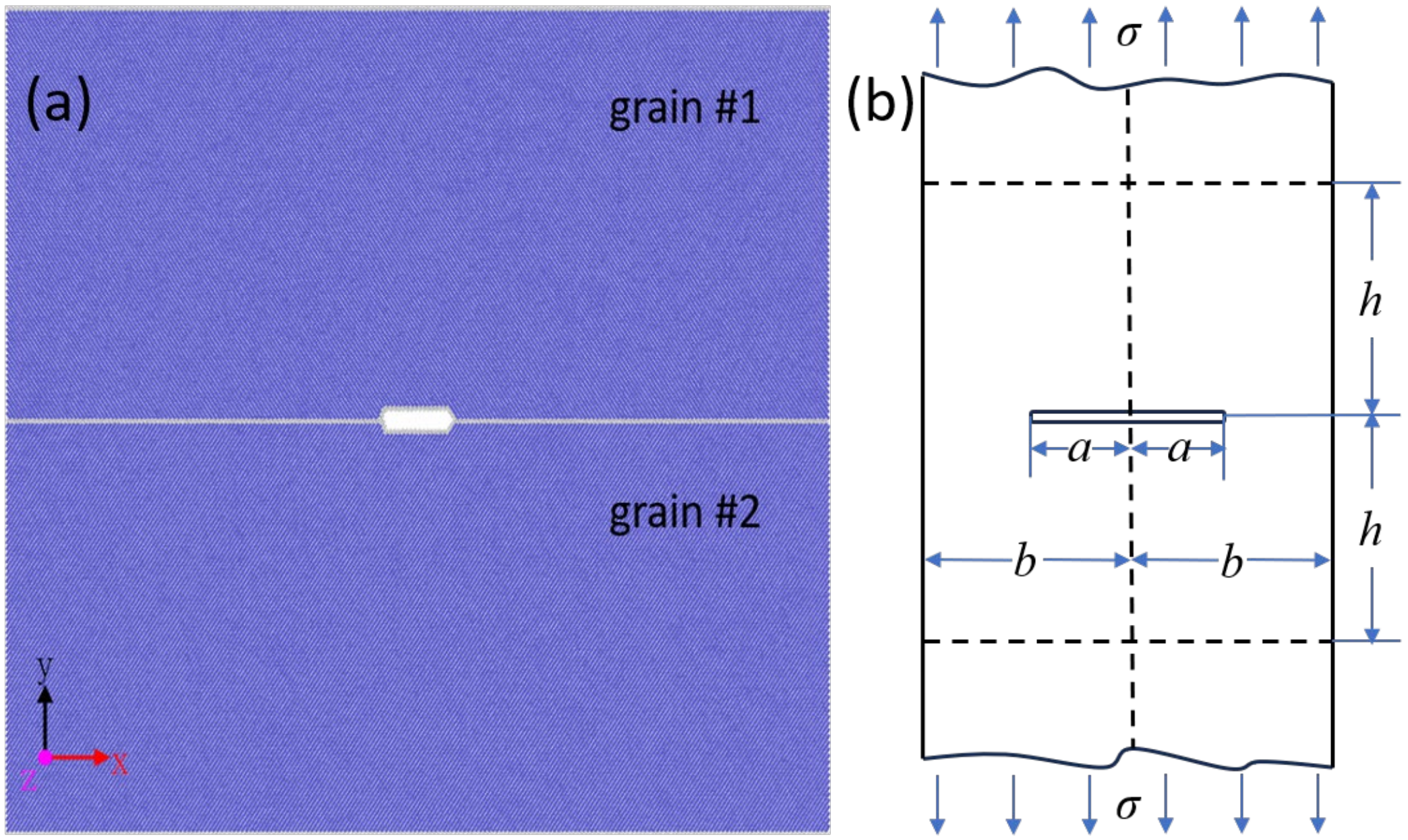

**Fig. 1**. (a) A typical bicrystalline specimen with a through-thickness crack at the GB. Periodic boundary condition is applied along the *x*- and *z*-axis directions, while the nonperiodic one along the *y*-axis direction. The size of all bicrystalline specimens is set as ~ 50×45×8 nm. The atoms are colored by the a-CNA parameter, with the bcc atoms as blue and disordered as white. (b) Schematics of the centered cacked test specimens.

## 3. Theoretical framework

While theoretical descriptions about the ductile blunting induced by the dislocation emission and the brittle cleavage proceeding by the breaking of atomic bonds have been well established for ideal lattice, counterparts for intergranular fracture are still under development. Subsequently, we would briefly review the recent rationalization of the GB cleavage in **Section 3.1**, then consider the Shimokawa-Tsuboi modification of the Rice-Thomson model in the classical Peierls framework to describe the GB sites as nucleation sources of dislocations near the crack-tip in **Section 3.2**. Furthermore, by coupling with the transition state theory, the rate and temperature dependencies are involved in the present framework appropriately.

### 3.1 The brittle cleavage along a GB

Within the Griffith model of the linear elastic fracture mechanics [52, 53], the critical SIF $K_{IG}$ for the brittle cleavage under Mode-I loading and plane strain conditions in the absence of plasticity is given by [54],

$$K_{IG} = \sqrt{\frac{G_I}{B}}, \tag{4}$$

where, $G_I$ is the energy release rate due to the formation of two new surfaces, and $B$ is the appropriate orientation dependent compliance constant [54, 55]. For anisotropic cubic crystals under plane strain conditions, $B$ can be written as,

$$B = \sqrt{\frac{b_{11}b_{22}}{2}\left(\sqrt{\frac{b_{22}}{b_{11}}} + \frac{2b_{12}+b_{66}}{2b_{11}}\right)}, \tag{5}$$

where, the plane strain moduli $b_{ij}$ can be further expressed in terms of elastic compliance constants $s_{ij}$ along the orientation of interest as [54-56],

$$\begin{cases} b_{11} = \frac{s_{11}s_{33}-s_{13}^2}{s_{33}} \\ b_{22} = \frac{s_{22}s_{33}-s_{23}^2}{s_{33}} \\ b_{12} = \frac{s_{12}s_{33}-s_{13}s_{23}}{s_{33}} \\ b_{66} = \frac{s_{66}s_{33}-s_{26}^2}{s_{33}} \end{cases}. \tag{6}$$

The calculated $B$ values for each orientation are shown in **Table B1** in **Appendix B**. For the complete brittle fracture in an ideally single-crystal (*i.e*. with the absence of GBs), the critical energy release rate is equal to surface energy (per unit area) of two newly-created surface,

$$G_I^{sc} = 2\gamma_{surf} \, . \tag{7}$$

However, if the crack propagates along a GB between two grains 1 and 2 (in **Fig. 1**a), the effect of GB energy $\gamma_{gb}$ has to be considered to calculate the energy required to create two new surfaces, which could be crystallographically different [57, 58],

$$G_I^{gb} = \gamma_{surf}^1 + \gamma_{surf}^2 - \gamma_{gb} \, , \tag{8}$$

where, $\gamma_{surf}^1$ and $\gamma_{surf}^2$ are the surface energies of the adjoining grains 1 and 2, respectively. In the case of negligible plastic deformation, the fracture along symmetrical GBs should always be more favorable than cleavage (inside the grain) along a plane parallel to the GB ($\gamma_{surf}^1 = \gamma_{surf}^2 = \gamma_{surf}$) [8]. The calculated $G_I^{gb}$ and $K_{IG}$ are listed in **Table C1** in **Appendix C**. Since the surface cannot form continuously but only by the breaking of discrete atomic bonds, the discreteness of the crystal lattice manifests itself in the so-called lattice trapping [53, 59-65]. Specifically, it results in an atomically sharp crack to remain stable during loading upon an upper limiting value $K_+$, higher than the Griffith SIF $K_{IG}$. Likewise, during unloading, the crack position remains unchanged until $K_- < K_{IG}$ is reached. The lattice trapping range, $\Delta K$, is then defined as [8, 60, 65],

$$\Delta K = \frac{K_+}{K_-} - 1 \, , \tag{9}$$

with the analogue of lattice trapping for interfaces is commonly referred to as bond trapping, which includes the breaking of bonds in the structural units of GBs [8, 60].

### 3.2 The ductile blunting of a GB crack

By considering the local GB structural transformation after dislocation emission from a GB, Shimokawa and Tsuboi [5] proposed a modified ***Rice-Thomson*** model [16] as (denoted as the ***Rice-Thomson-Shimokawa-Tsuboi*** (***RTST***) model later),

$$f = f_K + f_{d^2} + f_{dd'} + f_{gbe} \, . \tag{10}$$

The first term $f_K$ in Eq.(10) can be directly calculated from the applied Mode-I SIF $K_I$, which is a function of the applied load $\sigma_A$ and the crack length. The stress components due to $K_I$ at the dislocation in the Cartesian coordinate system are given by,

$$\sigma_{xx} = \frac{K_I}{\sqrt{2\pi r}}\cos\frac{\theta}{2}\left(1 - \sin\frac{\theta}{2}\sin\frac{3\theta}{2}\right) - \sigma_A \ , \tag{11}$$

$$\sigma_{yy} = \frac{K_I}{\sqrt{2\pi r}}\cos\frac{\theta}{2}\left(1 + \sin\frac{\theta}{2}\sin\frac{3\theta}{2}\right) , \tag{12}$$

$$\tau_{xy} = \frac{K_I}{\sqrt{2\pi r}}\sin\frac{\theta}{2}\cos\frac{\theta}{2}\cos\frac{3\theta}{2} , \tag{13}$$

where, the second term $\sigma_A$ in the Eq.(11) represents the applied load, and is usually neglected [5], ($r$, $\theta$) is defined as the polar coordinate in **Fig. 2**. $f_K$ is then evaluated as $b\tau_{r\theta}$, where the shear component $\tau_{r\theta}$ in the polar coordinate system is calculated by the coordinate transformation as,

$$\tau_{r\theta} = \cos\theta\sin\theta\left(\sigma_{yy} - \sigma_{xx}\right) + (\cos^2\theta - \sin^2\theta)\tau_{xy} \ . \tag{14}$$

The second term $f_{d^2}$ in the Eq.(10) is the self-image force caused by the free surface effect. Initially, to distinguish the stress components $\sigma_{yy}$ and $\tau_{yx}$ along the crack surface, image dislocation distributions are introduced along the crack surface [66]. The image dislocation ($\zeta_i$) distribution at a distance $\beta$ from the crack-tip, caused by the dislocation $\zeta$ emitted from the GB (see **Fig. 2**), is given by,

$$F_x(\beta) = \frac{-b}{\pi\rho_i}\sqrt{\frac{r}{|\beta|}}\left\{\cos\eta\cos\left(\phi - \frac{\theta}{2}\right) + \frac{1}{2}\sin\theta\sin\left(\phi - \eta + \frac{\theta}{2}\right) - \sin\phi\sin\left(2\phi - \eta - \frac{\theta}{2}\right)\right\} , \tag{15}$$

and

$$F_y(\beta) = \frac{-b}{\pi\rho_i}\sqrt{\frac{r}{|\beta|}}\left\{2\sin\eta\cos\left(\phi - \frac{\theta}{2}\right) + \cos\left(2\phi - \frac{\theta}{2}\right)\sin(\phi - \eta) - \frac{1}{2}\sin\theta\sin\left(\phi - \eta + \frac{\theta}{2}\right)\right\} . \tag{16}$$

Here $F_x(\beta)d\beta$ represents the sum of the $x$-directional components of the Burgers vector of the image dislocations between $\beta$ and $\beta + d\beta$, $b$ is the magnitude of the Burgers vector, and $\phi$ and $\eta$ are angles defined in **Fig. 2**. Substituting the Eqs.(15) and (16) into the stress field equations of a discrete dislocation [67] and integrating the functions over $\beta$ along the crack surface, the stress fields imposed by all image dislocations are given by,

$$\begin{aligned}\sigma_{xx} = \frac{\mu}{2\pi(1-\upsilon)}\Bigg\{ &- \int_{-\infty}^{0} \frac{y[3(x-\beta)^2+y^2]}{[(x-\beta)^2+y^2]^2} F_x(\beta)\mathrm{d}\beta \\ &+ \int_{-\infty}^{0} \frac{(x-\beta)[(x-\beta)^2-y^2]}{[(x-\beta)^2+y^2]^2} F_y(\beta)\mathrm{d}\beta \Bigg\} ,\end{aligned} \tag{17}$$

$$\sigma_{yy} = \frac{\mu}{2\pi(1-\upsilon)}\left\{-\int_{-\infty}^{0}\frac{y[(x-\beta)^2-y^2]}{[(x-\beta)^2+y^2]^2}F_x(\beta)\mathrm{d}\beta + \int_{-\infty}^{0}\frac{(x-\beta)[(x-\beta)^2+3y^2]}{[(x-\beta)^2+y^2]^2}F_y(\beta)\mathrm{d}\beta\right\}, \tag{18}$$

and

$$\tau_{xy} = \frac{\mu}{2\pi(1-\upsilon)}\left\{-\int_{-\infty}^{0}\frac{(x-\beta)[(x-\beta)^2-y^2]}{[(x-\beta)^2+y^2]^2}F_x(\beta)\mathrm{d}\beta + \int_{-\infty}^{0}\frac{y[(x-\beta)^2-y^2]}{[(x-\beta)^2+y^2]^2}F_y(\beta)\mathrm{d}\beta\right\}. \tag{19}$$

Considering the practice of numerical solving, Shimokawa and Tsuboi [5] set the interval of the integration as -100 < $\beta$ < 0 nm, and calculated the $f_{d^2}$ in the same manner as $f_K$. Besides, some geometric constraints should be declared,

$$\eta = \arctan\left(\frac{r\sin\theta}{r\cos\theta-\alpha}\right), \tag{20}$$

$$\rho_r = \frac{r\sin\theta}{\sin\eta}, \tag{21}$$

$$\phi = \arctan\left(\frac{r\sin\theta}{r\cos\theta+\beta}\right), \tag{22}$$

and

$$\rho_i = \frac{r\sin\theta}{\sin\phi}, \tag{23}$$

where, the subscripts '$r$' and '$i$' in the Eqs.(21) and (23) represent real and imaginary (dislocations), respectively.

The third term $f_{dd'}$ in the Eq.(10) can be decomposed into the direct and indirect interaction force $f_{dd',d}$ and $f_{dd',i}$ caused by the residual dislocation $\zeta'$ after the dislocation $\zeta$ emission from the GB, as shown in **Fig. 2**. In other words, the residual dislocation $\zeta'$ exerts two types of stress fields to the dislocation $\zeta$: the direct and indirect stress fields. The direct force $f_{dd',d}$ can be evaluated as,

$$f_{dd',d} = -\frac{\mu b^2}{2\pi(1-\upsilon)\rho_r}. \tag{24}$$

On the other hand, the indirect force $f_{dd',i}$ can be calculated in the same manner as $f_{d^2}$, where the image dislocation distribution caused by the residual dislocation $\zeta'$ is obtained by setting $b=-b$, $\rho_i=\alpha+\beta$, $r = \alpha$, and $\phi = \eta = \theta = 0$. Then, substituting the distributions into the Eqs.(17)-(19), the value of $f_{dd',i}$ can be determined.

The fourth term $f_{gbe}$ in the Eq.(10) is hypothesized to be caused by the local GB structural transformation from GB1 to GB0 after dislocation emission from the GB, as shown in **Fig. 2**. The GB dislocation κ is decomposed into a lattice dislocation $\zeta$ and a new GB dislocation $\zeta''$ with the Burgers vector parallel to the GB plane. However, Shimokawa and Tsuboi [5] suggested that the influence of $\zeta''$ on $\Delta E_{gb}$ can be ignored since $\zeta''$ does not contribute to the misorientation angle change of GBs and the residual dislocation effect on $f$ is already considered in $f_{dd'}$. The energy change due to the dislocation emission is then given by,

$$\Delta E_{gb} = \Delta \gamma_{gb} h \,, \tag{25}$$

where, $\Delta \gamma_{gb}$ is the GB energy difference between GB1 and GB0, and $h$ is the spacing between GB dislocations. The GB-induced force $f_{gbe}$ can be calculated analogously to the surface-step force [16] as,

$$f_{gbe} = -\frac{4\Delta E_{gb}\rho_0}{\pi(\rho_0^2+\rho_r^2)} \,, \tag{26}$$

where, the cutoff distance $\rho_0$ can be determined by comparing the two critical SIFs of dislocation emission from the crack-tip, *i.e.*, the Rice-Thomson model based on the elasticity solutions for a fully formed dislocation [16] and the Rice approach based on the Peierls concept [12], as follows,

$$\rho_0 = \frac{1-\upsilon}{2\pi\mu\gamma_{usf}}\left(\frac{\mu b}{2\sqrt{2}(1-\upsilon)} + 2\sqrt{2}\gamma_{surf}\right)^2 \frac{1}{1+(1-\upsilon)\tan^2\chi} \,, \tag{27}$$

where, $\gamma_{usf}$ and $\chi$ represent the unstable stacking fault energy and the angle of the slip direction on the slip plane, respectively.

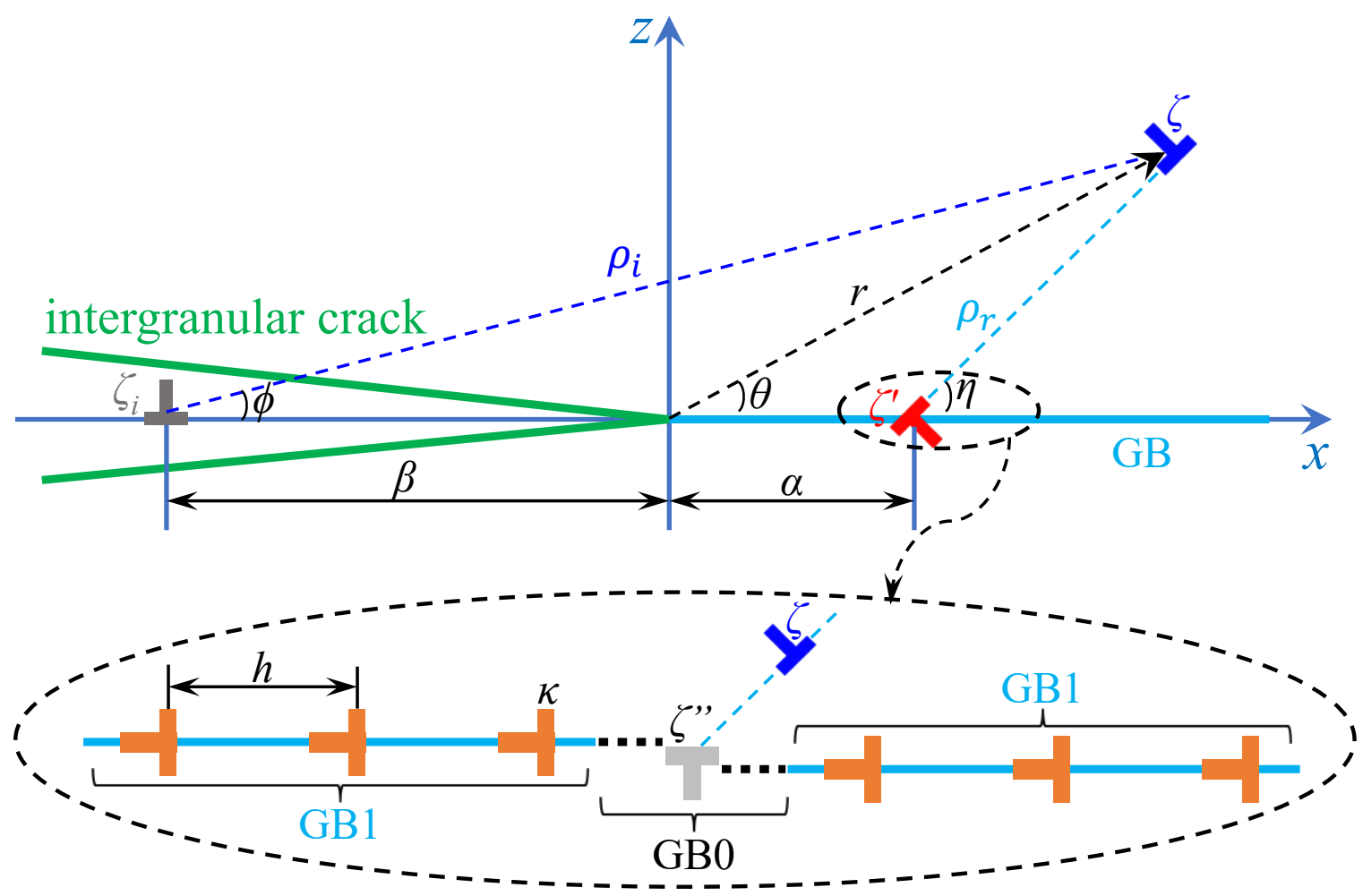

**Fig. 2**. Schematics of the analysis model by considering the dislocation emission from the GB ahead of the intergranular crack-tip proposed by Shimokawa and Tsuboi [5]. Change of the localized GB structure in GB0 is caused by a dislocation $\zeta$ emitted from GB1, composed of GB dislocations $\kappa$ with $b_\kappa = b_\zeta + b_{\zeta''}$.

To understand the origins of temperature-dependent activation energy barriers for dislocation nucleation from the crack-tip, a 2D Peierls model has been developed by Rice [12] and Rice and Beltz [13] as (denoted as the ***Peierls-Rice-Beltz*** (***PRB***) framework later),

$$U[\vec{\delta}(r)] = U_0 + \int_0^\infty \Phi[\vec{\delta}(r)]\mathrm{d}r + \frac{1}{2}\int_0^\infty s[\delta(r)] \cdot \delta(r)\mathrm{d}r - \int_0^\infty \frac{K_{II}^{eff}}{\sqrt{2\pi r}}\delta(r)\mathrm{d}r \,, \tag{28}$$

with

$$s[\delta(r)] = \frac{\mu}{2\pi(1-\upsilon)}\int_0^\infty \sqrt{\frac{\xi}{r}}\frac{\mathrm{d}\delta(\xi)/\mathrm{d}\xi}{r-\xi}\mathrm{d}\xi \,, \tag{29}$$

where, the first term $U_0$ is the elastic strain energy of the loaded cracked solid without any slip; the potential $\Phi$ is the change in atomic stacking energy due to a slip discontinuity $\vec{\delta}$; the third term accounts for the elastic interaction energy between the infinitesimal increments of slip; the fourth term represents the elastic interaction energy between the slip and the crack surface.

The activation energy per unit length under applied loading is the difference between the energies for the stable equilibrium slip distribution $\vec{\delta}_{stable}(r)$ and the saddle-point slip distribution $\vec{\delta}_{saddle}(r)$,

$$Q_{2d} = U[\vec{\delta}_{saddle}(r)] - U[\vec{\delta}_{stable}(r)] \,. \tag{30}$$

Proceeding via a perturbation analysis of the shear distribution, Rice and Beltz [13] obtained a closed-form approximation of the activation energy as,

$$\Theta_{2d} = \frac{(1-\upsilon)Q_{2d}}{\mu b^2} = m\left(1-\sqrt{\frac{G}{\gamma_{usf}}}\right)^{3/2} , \tag{31}$$

where, the dimensionless factor $m$ is approximated as 0.287 due to the extremely weak dependence on $\gamma_{usf}/\mu b$ [68, 69], the applied energy release rate $G$ can be correlated with the effective Mode-II SIF $K_{II}^{eff}$ via $G = (1-\upsilon){K_{II}^{eff}}^2/2\mu$ [69].

By defining a localized Mode-I SIF $K_I^*$ nominally, we can rewrite the total force in the ***RTST*** model as,

$$f = b\tau_{r\theta}^* , \tag{32}$$

where, $\tau_{r\theta}^*$ is the nominal shear component within the polar coordinate system, and can be related to the stress components within the Cartesian coordinate system via,

$$\tau_{r\theta}^* = \cos\theta\sin\theta\left(\sigma_{yy}^* - \sigma_{xx}^*\right) + (\cos^2\theta - \sin^2\theta)\tau_{xy}^* , \tag{33}$$

with the nominal stress components in the Cartesian coordinate as,

$$\sigma_{xx}^* = \frac{K_I^*}{\sqrt{2\pi r}}\cos\frac{\theta}{2}\left(1 - \sin\frac{\theta}{2}\sin\frac{3\theta}{2}\right) , \tag{34}$$

$$\sigma_{yy}^* = \frac{K_I^*}{\sqrt{2\pi r}}\cos\frac{\theta}{2}\left(1 + \sin\frac{\theta}{2}\sin\frac{3\theta}{2}\right) , \tag{35}$$

$$\tau_{xy}^* = \frac{K_I^*}{\sqrt{2\pi r}}\sin\frac{\theta}{2}\cos\frac{\theta}{2}\cos\frac{3\theta}{2} . \tag{36}$$

The shear stress can be rewritten as,

$$\tau_{r\theta}^* = \frac{K_I^*}{2\sqrt{2\pi r}}\sin\theta\cos\frac{\theta}{2} . \tag{37}$$

Thus,

$$K_I^* = \frac{2\sqrt{2\pi r}f}{b\sin\theta\cos\frac{\theta}{2}} . \tag{38}$$

To involve the effect of GBs on the dislocation nucleation, here we relate the effective Mode-II SIF $K_{II}^{eff}$ in Eq.(31) with the above-defined localized Mode-I SIF $K_I^*$ in Eq.(38) of the ***RTST*** model,

$$K_{II}^{eff} = K_I^* \cos^2\frac{\theta}{2}\sin\frac{\theta}{2}\,, \tag{39}$$

*i.e.*,

$$K_{II}^{eff} = \frac{\sqrt{2\pi r}f}{b}\,. \tag{40}$$

Thus, the ***PRB*** framework now can be extended to consider the dislocation emission from a GB crack via,

$$\frac{(1-\upsilon)Q_{2d}}{\mu b^2} = m\left(1-\sqrt{\frac{1-\upsilon}{\mu\gamma_{usf}}}\frac{\sqrt{\pi r}f}{b}\right)^{3/2}\,. \tag{41}$$

As seen from the Eq.(41), the condition for the existence of real roots is,

$$f \le \frac{\sqrt{\mu\gamma_{usf}}b}{\sqrt{(1-\upsilon)\pi r}}\,, \tag{42}$$

*i.e.*, the nominal Mode-I SIF needs to satisfy,

$$K_I^* \le \frac{2\sqrt{2\mu\gamma_{usf}}}{\sqrt{1-\upsilon}\sin\theta\cos\frac{\theta}{2}}\,. \tag{43}$$

The above activation energy derived as an implicit function of the applied Mode-I SIF $K_I$, however, cannot be directly employed in the transition-state-theory-based analysis of dislocation nucleation in following steps. Inspired by previous studies [30, 70], and considering the temperature dependence, the activation Gibbs free energy can be rewritten as,

$$Q_{2d}(K_I,T) = (1-T/T_m)Q_{2d}(K_I)\,, \tag{44}$$

where, $T_m$ is the surface disordering temperature (which to a first approximation, can be taken as the melting temperature) [70], and the activation enthalpy under zero temperature $Q_{2d}(K_I,T=0)$ is given by,

$$Q_{2d}(K_I) = C\left(1-\frac{K_I}{K_I^0}\right)^n\,, \tag{45}$$

where, $C$, $n$ and $K_I^0$ are fitting parameters obtained by fitting the zero temperature activation enthalpy $Q_{2d}$ values calculated from Eq.(41) to Eq.(45). In practice, we use the normalized form of Eq.(45), *i.e.*, $\Theta_{2d}(K_I) = \tilde{C}\left(1-\frac{K_I}{K_I^0}\right)^n$ with the dimensionless parameter $\tilde{C} = C\frac{1-\upsilon}{\mu b^2}$. The average rate of the dislocation nucleation can be described as,

$$\omega = \omega_0 N \exp\left(-\frac{Q_{3d}(K_I,T)}{k_B T}\right), \tag{46}$$

where, $\omega_0$ is the attempt frequency (to a first approximation, can be estimated as the Debye frequency, *i.e.*, $k_B T_D/\hbar$, where $T_D$ is the Debye temperature). Under the ***transition state theory*** (***TST***) framework, a dislocation nucleation event will occur once the applied Mode-I SIF $K_I$ is equal to the most probable SIF $K_{Id}^p$,

$$\left.\frac{Q_{3d}(K_I,T)}{k_B T}\right|_{K_I=K_{Id}^p} = \ln\left.\left(\frac{k_B T N \omega_0}{\dot{K}_I \Omega(K_I,T)}\right)\right|_{K_I=K_{Id}^p}, \tag{47}$$

where, $N\omega_0$ represents the number of potential nucleation sites in the vicinity of the crack-tip multiplied by the attempt frequency $\omega_0$, which will be treated as the Debye frequency ($\omega_0 = k_B T_D/\hbar$, where is the Debye temperature at the room temperature) here [71]. The 3D energy barrier $Q_{3d}$ is estimated from the 2D energy barrier $Q_{2d}$ with a scaling factor $s_0$, *i.e.*, $Q_{3d} = s_0 Q_{2d}$. The activation volume-like term is defined as,

$$\Omega(K_I,T) = -\frac{\partial Q_{3d}}{\partial K_I} = \left(1 - \frac{T}{T_m}\right)\frac{s_0 C n}{K_I^0}\left(1 - \frac{K_I}{K_I^0}\right)^{n-1}. \tag{48}$$

By numerically solving the Eq.(47) with the material constants listed in **Table 2**, the most probable SIF $K_{Id}^p$ for dislocation nucleation from a GB crack can be obtained. Comparing the values of $K_{Id}^p$ and $K_{IG}$, the fracture patterns can be determined as, *i.e.*, either the brittle cleavage or the ductile blunting via dislocation emission. The total framework is schematically concluded in **Fig. 3**.

**Table 2**. Material parameters of bcc Fe.

|  |  |  |
| --- | --- | --- |
| Burgers vector, $b$ (Å) | 2.4825 | Ref.[13] |
| shear modulus, $\mu$ (GPa) | 69.3 | Ref.[13] |
| Poisson's ratio, $\upsilon$ | 0.291 | Ref.[13] |
| surface energy, $\gamma_{surf}$ (J/m$^2$) | 2.37 | Ref.[12] |
| $\gamma_{surf}/\gamma_{usf}$ | 2.2 (Frenkel) | Ref.[12] |
| | 3.2 (EAM) | |
| Debye temperature, $T_D$ (K) | 477 (at 0 K) | Ref.[72] |
| | 373 (at 298 K) | Ref.[73] |
| surface disordering temperature, $T_m$ (K) | 1811 | Ref.[74] |

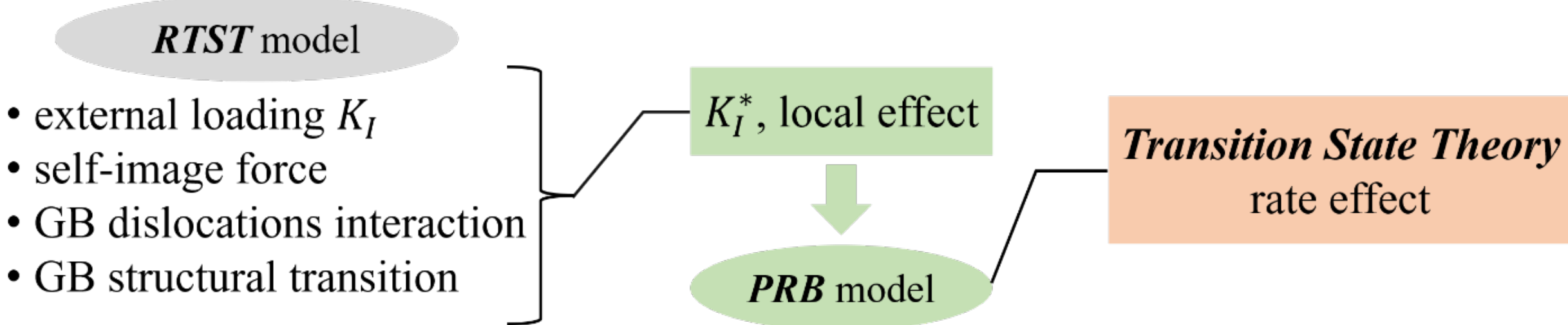

**Fig. 3**. Schematics of the theoretical framework constructed in this work. Here, ***RTST*** and ***PRB*** are abbreviations of Rice-Thomson-Shimokawa-Tsuboi and Peierls-Rice-Beltz, respectively.

## 4. Results

This section presents detailed analysis of MD simulation results, beginning with the global stress evolutions of bicrystalline specimens under uniaxial tension. Subsequently, the atomic deformation patterns near the crack tip are described in details. Additionally, the evolution of Mode-I SIF is analyzed by extracting the stress profile along the x-axis direction.

### 4.1 Stress evolution under uniaxial tension

In order to simulate the intergranular fracture behavior, the bicrystalline specimens are uniaxially stretched along the $y$-axis direction. A typical stress-strain curve is shown in **Fig. 4**a. Then, we extract the maximum values ($\sigma^{max}$) along the stress-strain curve for all 12 specimens, and plot them as a function of the GB tilt angle in **Fig. 4**b. By observing the snapshots inserted in **Fig. 4**a, it is found that critical events (either the plastic initiation or brittle cleavage) occur near the crack-tip at the moment that the stress-strain curve reaches $\sigma^{max}$, which thus could be defined as the critical stress $\sigma_{cr}$. **Fig. 4**b shows that the critical stress $\sigma_{cr}$ (*i.e.* the maximum tensile stress $\sigma^{max}$ on the stress-strain curve) varies drastically from each other, and no monotonic relation between the critical stress $\sigma_{cr}$ and GB characters can be derived.

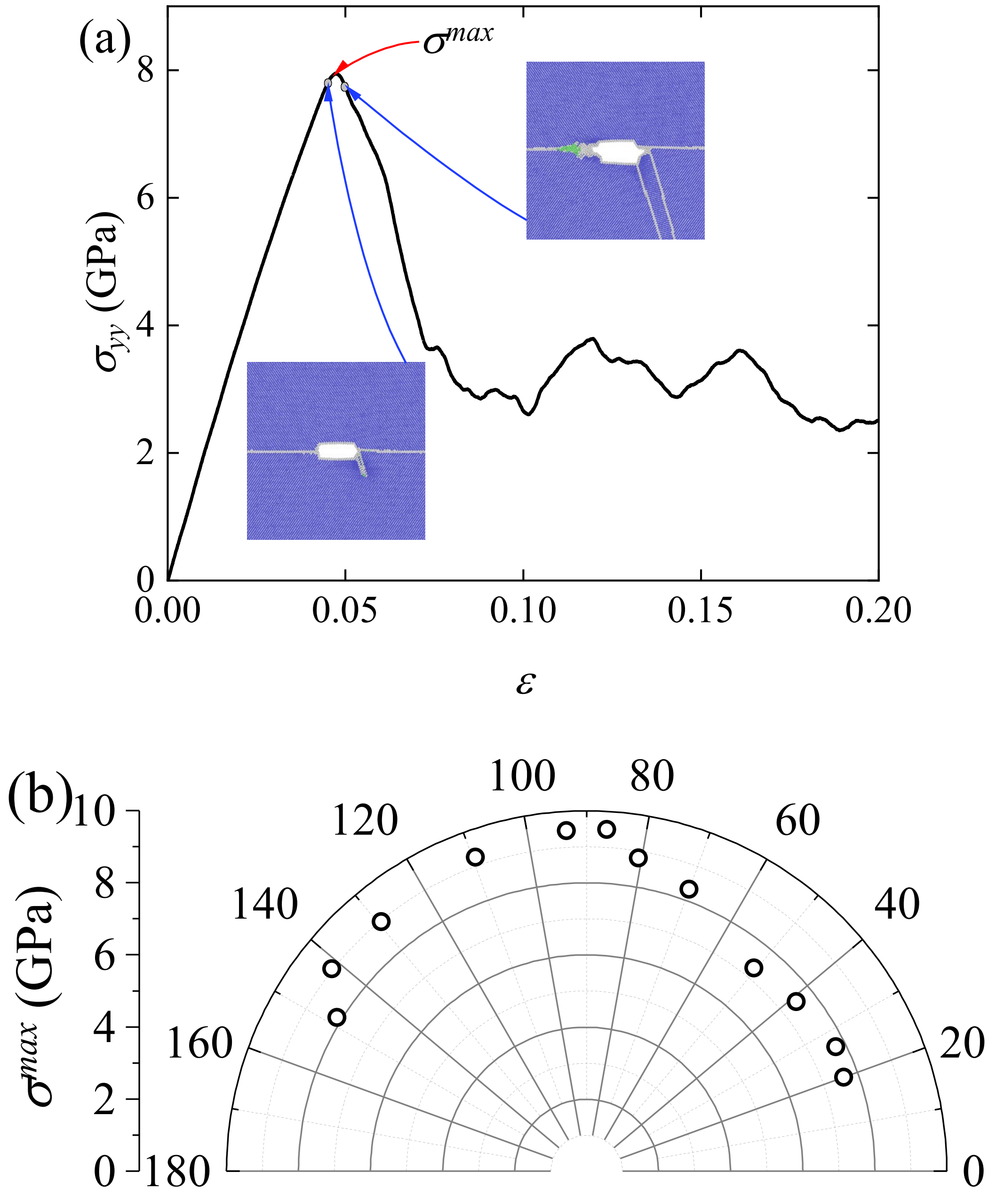

**Fig. 4**. (a) A typical stress-strain curve under uniaxial tension, with the maximum value $\sigma^{max}$ marked by the red arrow. The inserted panels show that critical events occur at the moment the stress-strain curve reaches the maximum $\sigma^{max}$, which thus could be defined as the critical stress $\sigma_{cr}$; (b) The maximum tensile stress $\sigma^{max}$ as a function of the GB tilt angle $\theta$ for all 12 bicrystalline specimens at 300 K (the strain rate $\dot{\varepsilon}$ = 5×10$^8$ /s).

Revisiting the Eq.(2), the Mode-I SIF $K_I^{max}$ corresponding to the maximum stress $\sigma^{max}$ can be evaluated in **Fig. 5**, while the maximum stress in present work generally corresponds to the initiation of plastic events, as shown in **Fig. 6-8**. It is noted that the stress filed near the crack-tip in atomic models should be corrected as $F_1F_2K_I$, where $F_1$ and $F_2$ are the geometrical correlation factor due to the geometry of finite size specimens and an additional correlation factor (introduced to fit the results of atomistic simulations) due to the simple tensile boundary condition, respectively [5]. Zhang *et al*. [75] recently reported the $K_{Ie}$ (for dislocation emission) calculated by the MEAM potential is about 1.1 MPa$\sqrt{\text{m}}$, which is in reasonable agreement with the present results, if we consider the difference of $\gamma_{usf}$ due to the employed force fields (see **Appendix D**). The results also show that the critical SIFs required for dislocation emission (marked by the violet circles) predicted by the Rice model (see **Appendix D**) agree well with the simulated $K_I^{max}$ of single crystal (marked by the orange and olive open triangles) with middle rotation angles (*i.e.* the tilt angle $\theta$ of the corresponding bicrystalline specimens, about 70 ~ 110°), but deviate significantly from simulation results for both the lower (about 20 ~ 50°) and higher (about 130 ~ 150°) angles.

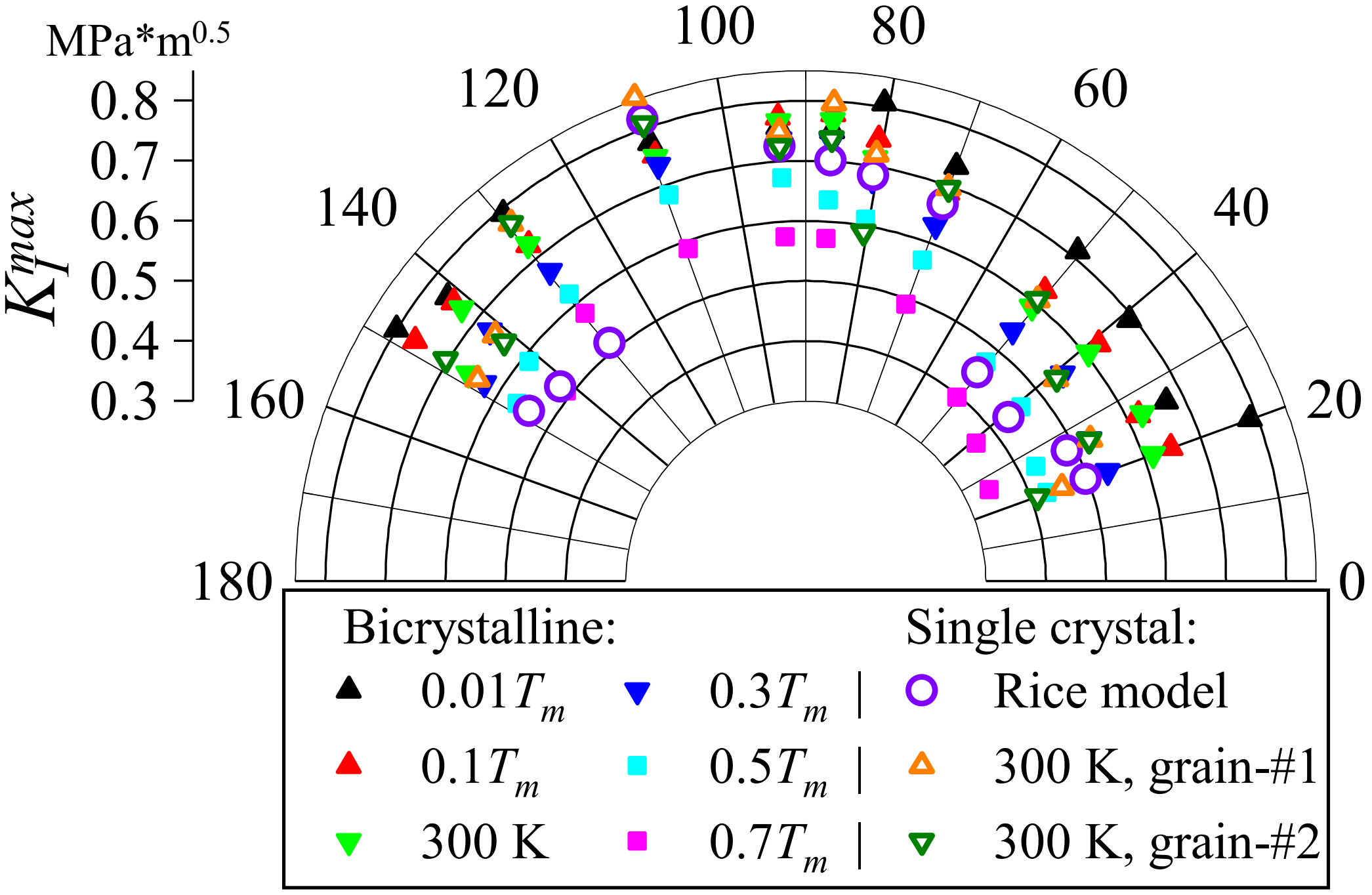

**Fig. 5**. The maximum Mode-I SIF $K_I^{max}$ evaluated from the maximum stress $\sigma^{max}$ along the simulated stress-strain curve via the Eq.(1) for tests at various temperature. The solid symbols represent the results of bicrystalline specimens. The open circles are critical SIF required for dislocation emission in single crystals evaluated by the classical Rice model considering the lattice anisotropy (see details in **Appendix A**, **B**, **C** & **D**). The open triangles represent the simulation results of single crystals with the lattice orientations shown in **Table 1**.

## 4.2 Deformation patterns near the crack-tip

**Fig. 6-8** show the evolution of crack propagation in these 12 bi-crystalline samples up to the tensile strain $\varepsilon$ = 6.0%. It is found that, 1) for the tilt angle $\theta \leq$ 80.60°, considerable twin activities are observed near the crack-tip; 2) for 86.60° $\leq \theta \leq$ 93.40°, twins start to nucleate at ~ $\varepsilon$ = 6.0%; 3) for $\theta \geq$ 109.5°, only dislocations are nucleated and emitted from the crack-tip or GB. It is interesting to find that the increase of $\sigma_{yy}^{max}$ (shown in **Fig. 4**) corresponds to the transition from twin nucleation to dislocation emission. It should be noted that the slight difference among the absolute values of $\sigma^{max}$ might not be interesting since the thermal fluctuation cannot be ignored under finite temperature. In other words, the simulated stress-strain curve of the same configuration might change slightly with the same simulation settings but only changing the initial distribution of velocities with a different random number.

Specifically, the deformation patterns of these specimens upon tensile loading are identified as, 1) Σ3(70.00): at $\varepsilon$ = 4.5%, a nanotwin was nucleated in the lower grain from the right crack-tip; at $\varepsilon$ = 5.0%, the left crack starts to propagate, together with the amorphization of the GB structure in front of the left crack-tip; at $\varepsilon$ = 5.5%, dislocation loops were emitted with one end attached on the GB plane and another on the newly-formed crack surface (which is originally also the GB plane). 2) Σ3(109.5): at $\varepsilon$ = 6.0%, a series of 1/2<111> dislocations are emitted from the newly-formed crack surfaces together with the crack propagation along the GB plane. 3) Σ9(38.90): at $\varepsilon$ = 5.5%, arrow-like twins were nucleated in the vicinity of the right crack-tip beside the GB; at $\varepsilon$ = 6.0%, a series of 1/2<111> dislocations are emitted from the interfaces between the original grains and newly-formed nanotwins. 4) Σ9(141.6): at $\varepsilon$ = 4.5%, a 1/2[$\bar{1}1\bar{1}$] dislocation was emitted into the upper grain from the GB in front of the right crack-tip, together with the amorphization of the dislocation source site at the GB; at $\varepsilon$ = 5.0%, another 1/2[$11\bar{1}$] dislocation was emitted into the lower grain from the GB again; at $\varepsilon$ = 6.0%, 1/2<111> dislocations were also emitted from the GB in front of the left crack-tip. 5) Σ11(50.50): at $\varepsilon$ = 4.5%, arrow-like twins nucleate beside the GB in front of the right crack, together with the distortion of the GB structure inside the newly-formed nanotwins; meanwhile, a series of 1/2<111> dislocations also nucleate on the interface between the nanotwin and the lower grain; at $\varepsilon$ = 5.0%, a 1/2[$\bar{1}1\bar{1}$] dislocation was emitted from the intersection of the crack surface and the twin boundary; at $\varepsilon$ = 5.5%, the crack propagates along the GB from the left crack-tip. 6) Σ11(129.5): at $\varepsilon$ = 4.5%, a 1/2[$\bar{1}1\bar{1}$] dislocation loop was emitted into the upper grain from the GB in front of the left crack-tip; at $\varepsilon$ = 5.0%, 1/2[$\bar{1}11$] dislocation loops were nucleated at the GB in front of the right crack-tip and a 1/2[$11\bar{1}$] dislocation was emitted into the lower grain from the GB in front of the left crack-tip. 7) Σ17(86.60): at $\varepsilon$ = 6.0%, the intergranular crack propagates from the left crack-tip, and twins were nucleated from the right crack-tip. 8) Σ17(93.40): at $\varepsilon$ = 6.0%, the intergranular crack propagates from the left crack-tip, and twins were nucleated from the right crack-tip. 9) Σ19(26.50): at $\varepsilon$ = 5.5%, a twin was emitted from the right crack-tip in the upper grain, meanwhile a 1/2[$\bar{1}1\bar{1}$] dislocation was nucleated in the GB in front of the right crack-tip. 10) Σ27(148.4): at $\varepsilon$ = 4.0%, 1/2<111> dislocations were emitted from the GB in front of the left crack-tip; at $\varepsilon$ = 5.5%, a void nucleates at the GB with a certain distance from the left crack-tip; at $\varepsilon$ = 6.0%, the void continuously grows along the GB, and develops into a new crack. 11) Σ33(20.05): at $\varepsilon$ = 5.5%, twins were nucleated from the intersection of the right crack-tip and GB; at $\varepsilon$ = 6.0%, a twin was nucleated from the GB in front of the left crack-tip, accompanied with the crack propagation. 12) Σ43(80.60): at $\varepsilon$ = 5.0%, a twin embryo was nucleated from the intersection of the right crack-tip and GB; at $\varepsilon$ = 5.5%, the intergranular crack propagates along the GB towards the negative $x$-direction, accompanied with the 1/2<111> dislocations nucleated on the newly-formed crack surface. For readers' convenience, we have summarized the fracture behaviors above in **Table 3**.

**Table 3**. Crack-tip behaviors around GBs. Here the abbreviations "dislo.", "nuclea.", "amorph." and "struct. trans." represent "dislocation", "nucleation", "amorphization", and "structural transition", respectively. The "R-tip" and "L-tip" represent the right crack-tip and left crack-tip, respectively.

| | $\varepsilon$ = 0.0% | $\varepsilon$ = 4.5% | $\varepsilon$ = 5.0% | $\varepsilon$ = 5.5% | $\varepsilon$ = 6.0% |
|---|---|---|---|---|---|
| Σ3(70.00) | --- | twin nuclea. at R-tip | amorph. at L-tip | dislo. loop near L-tip | --- |
| Σ3(109.5) | --- | --- | --- | --- | GB cleavage & twin dislocations |
| Σ9(38.90) | --- | arrow-like twin nuclea. at R-tip | --- | cleavage at L-tip | --- |
| Σ9(141.6) | --- | dislo. emission at R-tip | GB struct. trans. | --- | dislo. emission from GB site at L-tip |
| Σ11(50.50) | --- | arrow-like twin nuclea. at R-tip | --- | cleavage at L-tip | --- |
| Σ11(129.5) | --- | dislo. emission from GB site at L-tip | dislo. emission from GB site at R-tip | --- | GB struct. trans. |
| Σ17(86.60) | --- | --- | --- | --- | cleavage at L-tip; twin nuclea. at R-tip |
| Σ17(93.40) | --- | --- | --- | --- | cleavage at L-tip; twin nuclea. at R-tip |
| Σ19(26.50) | dislo.-array like small angle GB | --- | --- | twin nuclea. at R-tip | twin nuclea. at L-tip |
| Σ27(148.4) | --- | dislo. emission from GB site at L-tip | --- | void nuclea. at GB ahead of L-tip; GB struct. trans. ahead of R-tip | cleavage ahead of L-tip |
| Σ33(20.05) | dislo.-array like small angle GB | --- | --- | twin nuclea. at R-tip | twin nuclea. at L-tip |
| Σ43(80.60) | --- | --- | twin nuclea. at R-tip | cleavage at L-tip | crack propagation accompanied with amorph. |

In summary, consistent with the predictions of classical Rice model (see details in **Table C1** in **Appendix C**), it is found that no plasticity was activated before the crack propagation in the Σ3(109.5) specimen. However, intergranular cleavage was also found for the left-side crack of Σ9(38.90), Σ11(50.50), Σ17(86.60) and Σ17(93.40) specimens since the effect of GB structure anisotropy is not considered in the energy-based theory (see **Appendix C**).

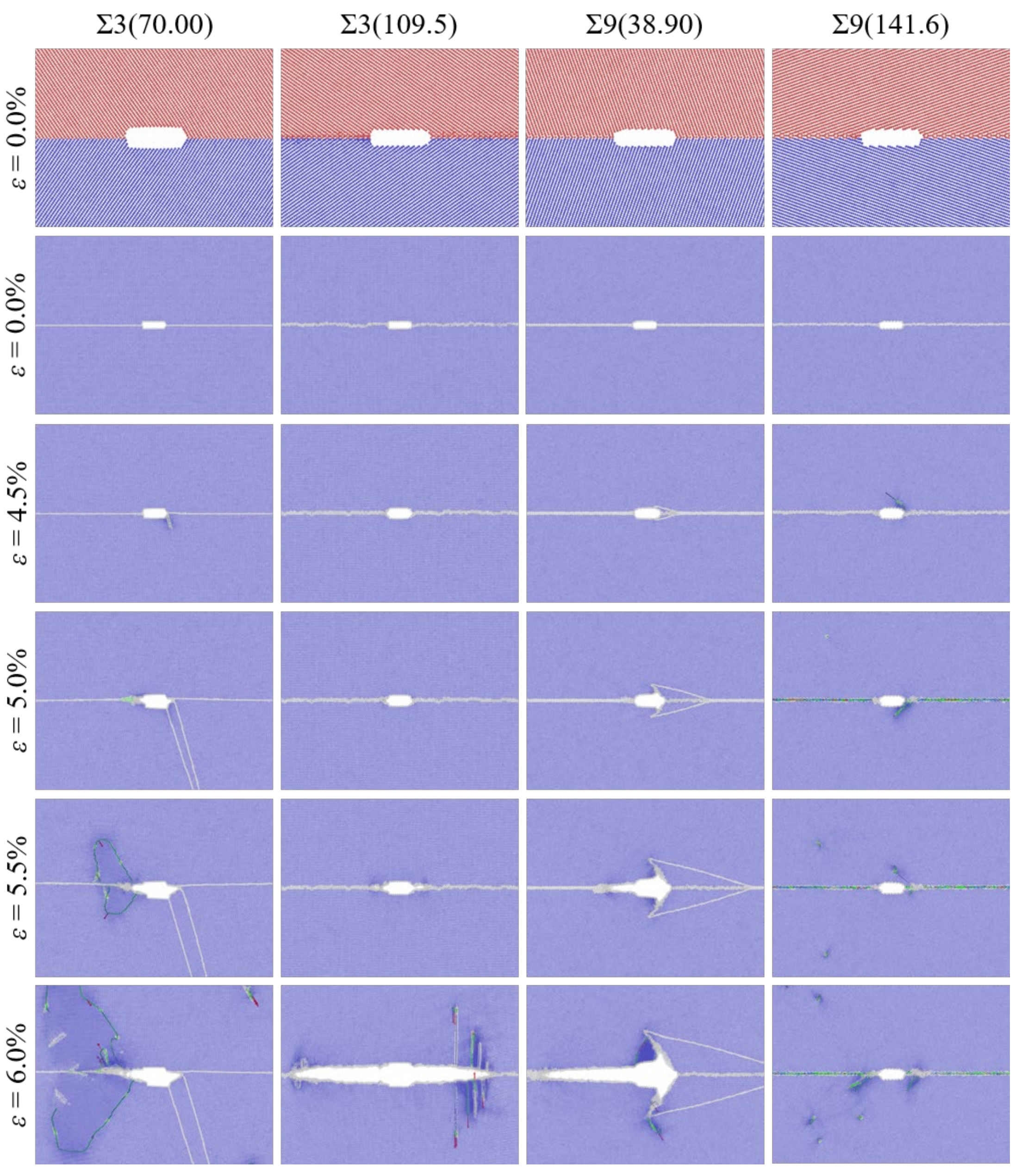

**Fig. 6**. Evolution of atomic configurations around the crack along the Σ3(70.00), Σ3(109.5), Σ9(38.90) and Σ9(141.6) GBs. The first row (*i.e.*, the samples are not loaded yet) panels are colored by the particle types, and enlarged for a better view. The other configurations under loading are color by the adaptive common neighbor analysis (a-CNA) processing, with the bcc, fcc, hcp and unidentified structures are marked as blue, green, red and white, respectively. It is noted that the first-row configurations are enlarged for a better view of the GB structural units.

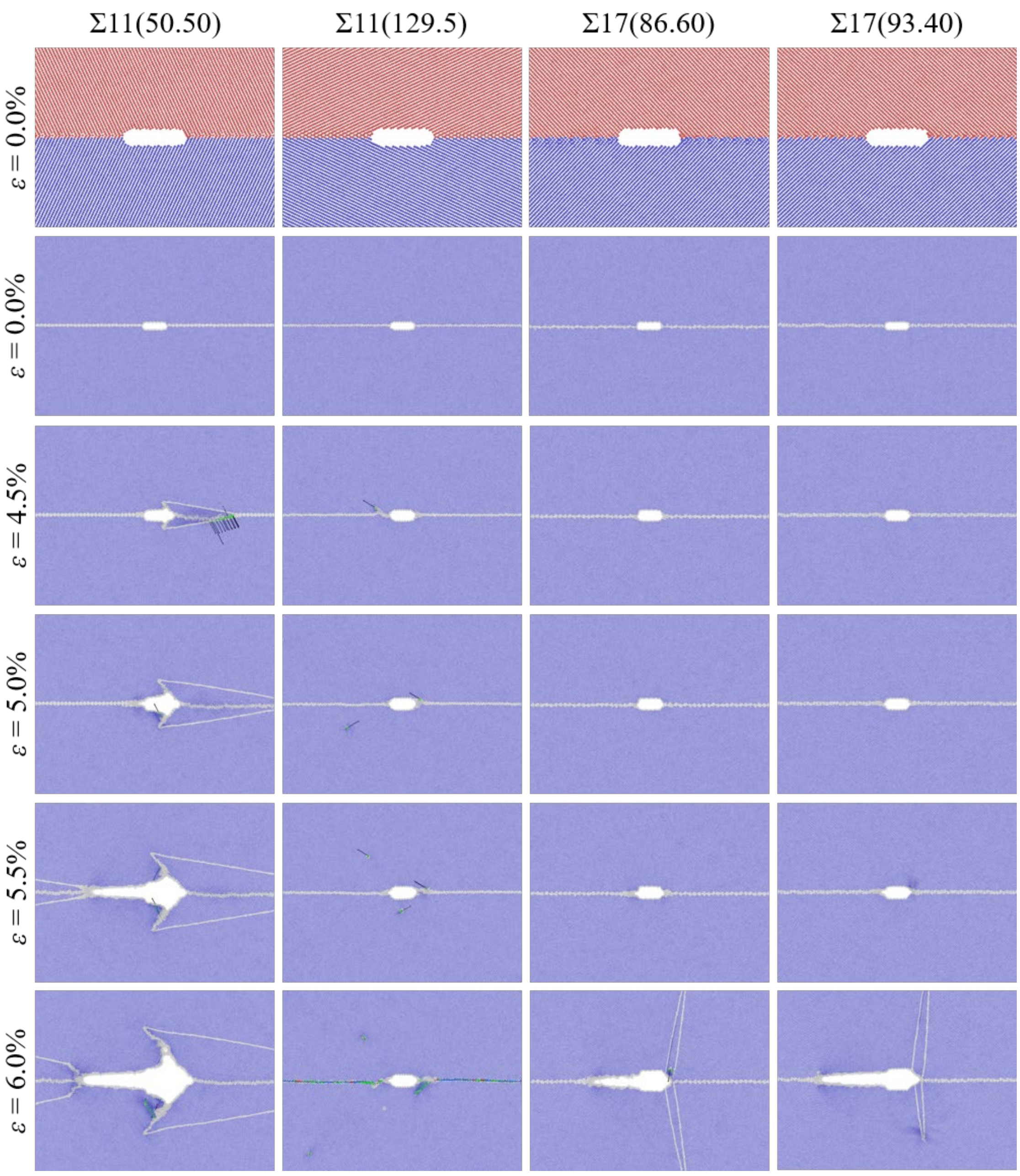

**Fig. 7**. Evolution of atomic configurations around the crack along the Σ11(50.50), Σ11(129.5), Σ17(86.60) and Σ17(93.40) GBs. The first row (*i.e.*, the samples are not loaded yet) panels are colored by the particle types, and enlarged for a better view. The other configurations under loading are color by the adaptive common neighbor analysis (a-CNA) processing, with the bcc, fcc, hcp and unidentified structures are marked as blue, green, red and white, respectively. It is noted that the first-row configurations are enlarged for a better view of the GB structural units.

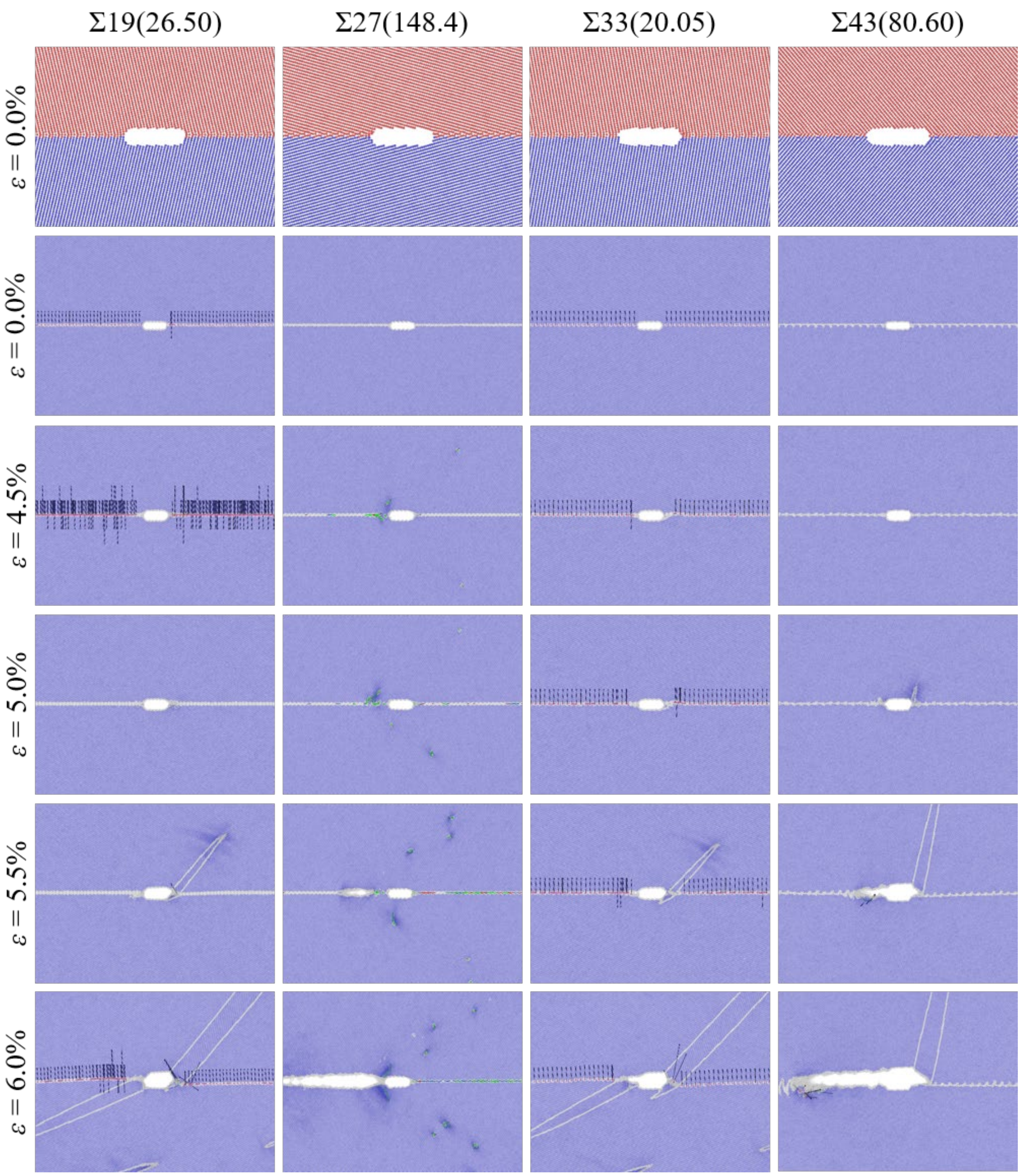

**Fig. 8**. Evolution of atomic configurations around the crack along the Σ19(26.50), Σ27(148.4), Σ33(20.05) and Σ43(80.60) GBs. The first row (*i.e.*, the samples are not loaded yet) panels are colored by the particle types, enlarged for a better view. The other configurations under loading are color by the adaptive common neighbor analysis (a-CNA) processing, with the bcc, fcc, hcp and unidentified structures are marked as blue, green, red and white, respectively. It is noted that the first-row configurations are enlarged for a better view of the GB structural units.

**Table 4** summarizes the crack-tip behaviors of these 12 samples. Generally speaking, the propagation of the right-side crack was suppressed in all samples except the Σ3(109.5) sample, in which the emitted twin boundaries are perpendicular to the crack surfaces. As predicted by the classical theory reviewed in **Section 3.1** and **Appendix C**, the $K_{IG}$ of Σ3(109.5) GB is smaller than $K_{Ie}$, thus the brittle cleavage is favored. It is noted that the twin nucleation in the Σ9(38.90) and Σ11(50.50) samples leads to the formation of an arrow-shaped twinned region beside the GB plane. It is also found that the crack propagation could be the cleavage without any other plastic activities, or promoted by phase transition, nanovoiding and amorphization, while the crack-tip blunting is always accompanied with twin nucleation or dislocation emission.

**Table 4**. Intergranular fracture behavior of 12 bi-crystalline samples up to $\varepsilon$ = 6.0% with the environmental temperature $T$ = 300 K and loading rate $\dot{\varepsilon} = 5.0 \times 10^8$ /s, where the crack propagation occurs via either the intergranular (IG) fracture (marked by the left- or right-headed triangle to indicate the propagation direction), or is suppressed by the ductile blunting (DB, marked by the symbol ◙ to indicate the stopping of the crack propagation).

| GB index | left-side crack | | | right-side crack | | |
|---|---|---|---|---|---|---|
| | crack propagation | fract. type | crack-tip deform. pattern | crack propagation | fract. type | crack-tip deform. pattern |
| Σ3(70.00) | ◀ | DB | Disloc. slip | ◙ | DB | Twinning |
| Σ3(109.5) | ◀ | IG | - | ▶ | IG | - |
| Σ9(38.90) | ◀ | IG | - | ◙ | DB | Twinning |
| Σ9(141.6) | ◙ | DB | Disloc. slip | ◙ | DB | Disloc. slip |
| Σ11(50.50) | ◀ | IG | - | ◙ | DB | Twinning |
| Σ11(129.5) | ◙ | DB | Disloc. slip | ◙ | DB | Disloc. slip |
| Σ17(86.60) | ◀ | IG | - | ◙ | DB | Twinning |
| Σ17(93.40) | ◀ | IG | - | ◙ | DB | Twinning |
| Σ19(26.50) | ◙ | DB | Twinning | ◙ | DB | Twinning |
| Σ27(148.4) | ◀ | IG | Void | ◙ | DB | Disloc. slip |
| Σ33(20.05) | ◙ | DB | Twinning | ◙ | DB | Twinning |
| Σ43(80.60) | ◀ | IG | Amorph. | ◙ | DB | Twinning |

### 4.3 Evolution of the Mode-I SIF

**Fig. 9** shows the profile of the tensile stress $\sigma_{yy}$ along the $x$-axis corresponding to the atomistic configurations shown in **Fig. 6-8**. It is found that, 1) for the crack propagation with few dislocation activities, the local stress near the crack-tip could reach ~ 17.5 GPa and even larger, e.g. the Σ3(109.5), Σ9(38.90), Σ11(50.50), Σ17(86.60), Σ17(93.40), Σ43(80.60) samples; 2) the crack propagation accompanied by dislocation nucleation leads to the moderate stress level (~ 15 GPa) at the crack-tip, meanwhile the formation of plastic zone results in the localized concentration of the stress field, e.g. the Σ3(70.00) sample; 3) the twin nucleation reduces the stress level ahead of the crack-tip with the increasing strain, e.g. the Σ3(70.00), Σ9(38.90), Σ11(50.50), Σ17(86.60), Σ17(93.40), Σ19(26.50), Σ33(20.05), Σ43(80.60). It is noted that the local stress in the right-hand side of the Σ27(148.4) sample deceases with increasing strain, although no twins were nucleated. The atomic configuration shown in **Fig. 8** indicates that misfit dislocations nucleate at the Σ27(148.4) GB upon tensile straining. The local stress in Σ9(141.6) and Σ11(129.5) samples, by contrast, increases with the strain, accompanied by the nucleation of misfit dislocations and emission of dislocations from the GB. Besides, different from the Σ27(148.4) sample, an amorphous region forms ahead of the crack-tip in Σ9(141.6) and Σ11(129.5) samples.

In **Fig. 9**, regarding the position of the maximum tensile stress (upon which the local $\sigma_{yy}$ sharply decreases to almost zero) as the crack-tip in a first approximation, we can fit the $\sigma_{yy}$ distribution along the *x*-axis direction to the theoretical solution of fracture mechanics, *i.e.* $\sigma_{yy} = K_I(t)/\sqrt{2\pi r}$, via which the Mode-I dynamic SIF (DSIF) $K_I(t)$ in real time during the process of dynamic fracture can be obtained (fitting procedure is operated only for $r < 50\pm5$ Å for numerical stability). In order to determine the critical $K_I^c(t)$, at which the fracture or plasticity events occur, we plot the DSIF $K_I(t)$ as a function of the time $t$ in **Fig. 10**. It is found that before the occurrence of the critical events, the DSIF $K_I(t)$ generally increases linearly with the time $t$. For the Σ11(50.50) specimen, **Fig. 10** shows that the critical point, at which the DSIF $K_I(t)$ of the left crack deviates from the linearity, decreases with the increasing temperature. Meanwhile, it is found that the slope (*i.e.*, the SIF rate $\dot{K}_I$ in the unit $\mathrm{MPa}\sqrt{\mathrm{m}}/\mathrm{s}$) also decreases with the increasing temperature, although all simulations are performed under the same strain rate. However, compared with the $K_I^{max}$ evaluated from the global stress-strain curve shown in **Fig. 5**, critical values ($K_I^c(t)$) at which the DSIF $K_I(t)$ deviates from a linear relation are generally larger than the simulated $K_I^{max}$ values and classical Rice model predictions (see **Fig. 5**). This difference thus indicates that the effect of localized atomic configurations around the crack-tip has to be considered.

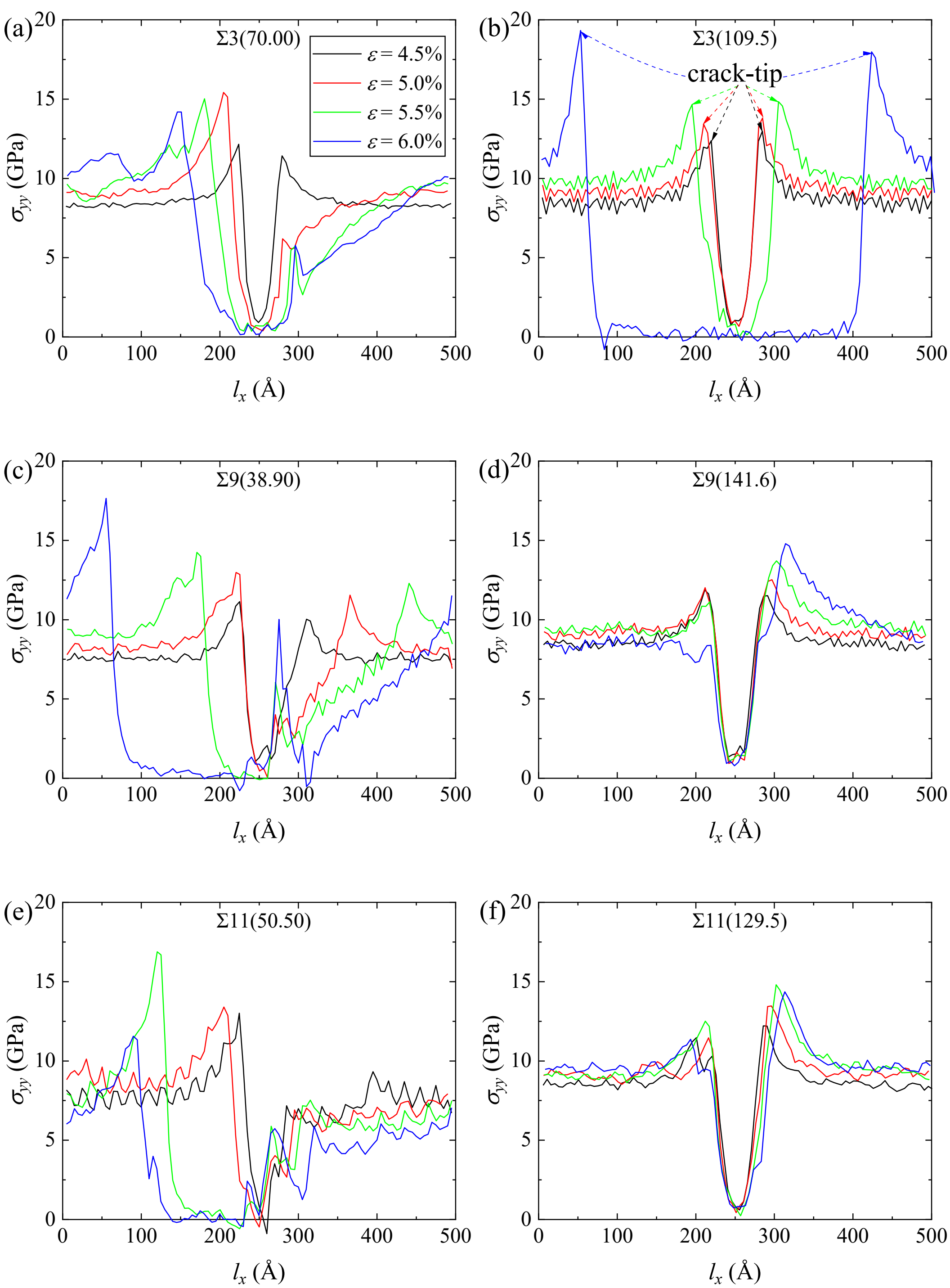

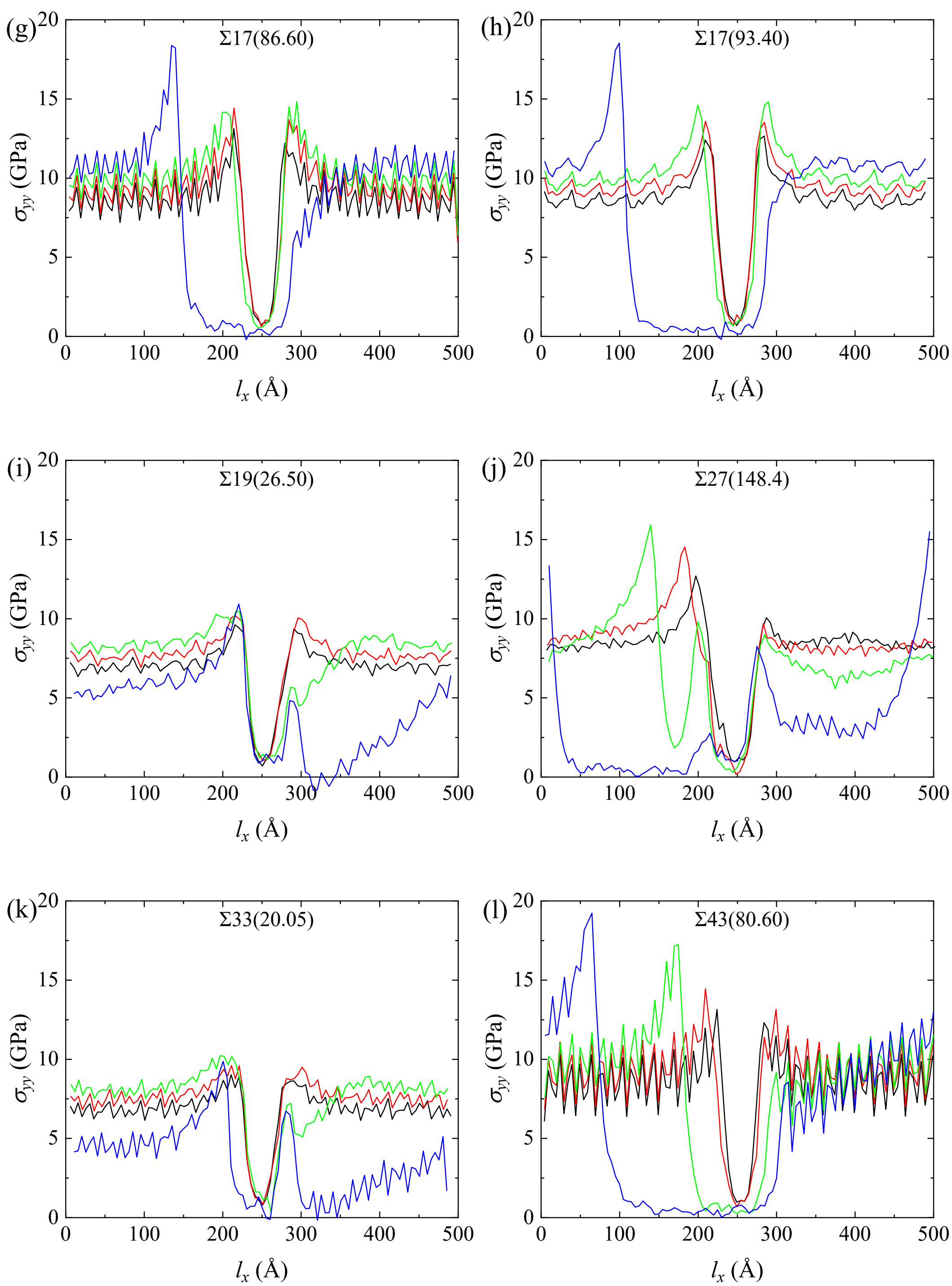

**Fig. 9**. Tensile stress profile at 300 K along the $x$-axis direction under different strains $\varepsilon = 4.5\%$, 5.0%, 5.5% and 6.0% for (a) Σ3(70.00); (b) Σ3(109.5); (c) Σ9(38.90); (d) Σ9(141.6); (e) Σ11(50.50); (f) Σ11(129.5); (g) Σ17(86.60); (h) Σ17(93.40); (i) Σ19(26.50); (j) Σ27(148.4); (k) Σ33(20.05); (l) Σ43(80.60) specimens, with the valley at the center of the $x$ dimension indicates the position of the crack. Meanwhile, as shown in panel (b), the local maximum near the center of the $x$-dimension indicates the crack-tip position. As demonstrated in **Appendix E**, the stress profile is outputted by cutting the rectangular region containing the GB into a series of bins. Thus, the sawtooth oscillation originates from the fact that $m \cdot l_b \approx n \cdot l_0$, where $m$ and $n$ are integers, $l_b$ is the bin size and $l_0$ is the lattice periodicity of the GB along the $x$-axis direction.

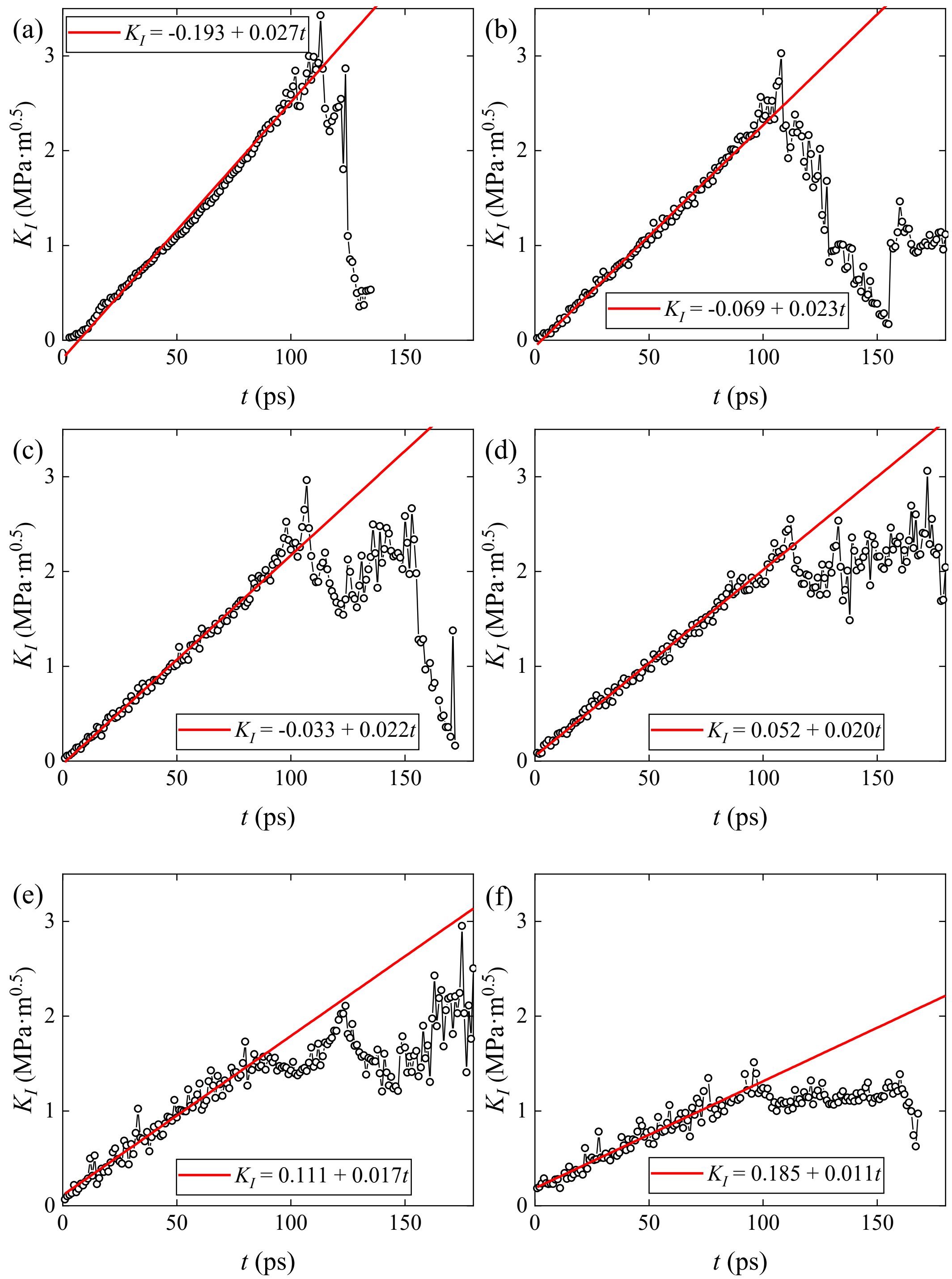

**Fig. 10**. The DSIF $K_I(t)$ as a function of time $t$ for the left crack of the Σ11(50.50) specimen at (a) $T = 0.01T_m$; (b) $T = 0.1T_m$; (c) $T = 300$ K; (d) $T = 0.3T_m$; (e) $T = 0.5T_m$; (f) $T = 0.7T_m$. Red lines are fitted to the MD results. It is noted

that the fitted slope, *i.e.* the DSIF rate $\dot{K}_I$, decreases with the increasing temperature while the specimen was loaded under the same strain rate.

Revisiting **Fig. 6-8**, it is found that the Σ3(109.5), Σ9(141.6), Σ11(129.5) and Σ27(148.4) specimens, which have the largest GB angles among the studied specimens, do not display any nano-twinning during the loading process. Thus, we can directly measure the DSIF of both the left and right cracks in these two specimens. At a given applied strain, the energy release rate scales linearly with the box size along the uniaxial loading direction (here the *y*-axis) as [26],

$$G_I = \frac{1}{2} C_{22} \varepsilon_{yy}^2 L_y \tag{49}$$

where, $C_{22}$ is the elastic modulus. According to the Griffith theory, the Eq.(49) can be rewritten as,

$$K_I = \sqrt{\frac{C_{22} L_y}{2B}} \varepsilon_{yy} \tag{50}$$

In a time-increment $\Delta t$,

$$\dot{K}_I \Delta t = \sqrt{\frac{C_{22} L_y}{2B}} \dot{\varepsilon}_{yy} \Delta t \tag{51}$$

Since all specimens are loaded with a constant strain rate uniaxially, the Mode-I SIF is anticipated to increase linearly with the time $t$ before the occurrence of critical events. The above Eq.(51) also explains why the $\dot{K}_I$ changes with the variation of temperature in **Fig. 10**, since the elastic modulus is anticipated to be a function of temperature.

**Fig. 11-14** show that the slope (marked by the solid fitting lines) of the left and right cracks in the Σ3(109.5), Σ9(141.6), Σ11(129.5) and Σ27(148.4) specimens does not differ from each other significantly before the occurrence of the critical events. However, it is found that the $K_I^c(t)$ values of the left and right crack might differ from each other, *i.e.* the critical events at the left and right crack-tips do not occur simultaneously under different temperatures. The temperature dependent anisotropy of the intergranular fracture behavior might be attributed to the geometrical asymmetry of the GB structural unit induced by the variation of temperature. These results suggest that the local variation of atomistic configuration near the crack-tip could drastically influence the initiation of crack propagation.

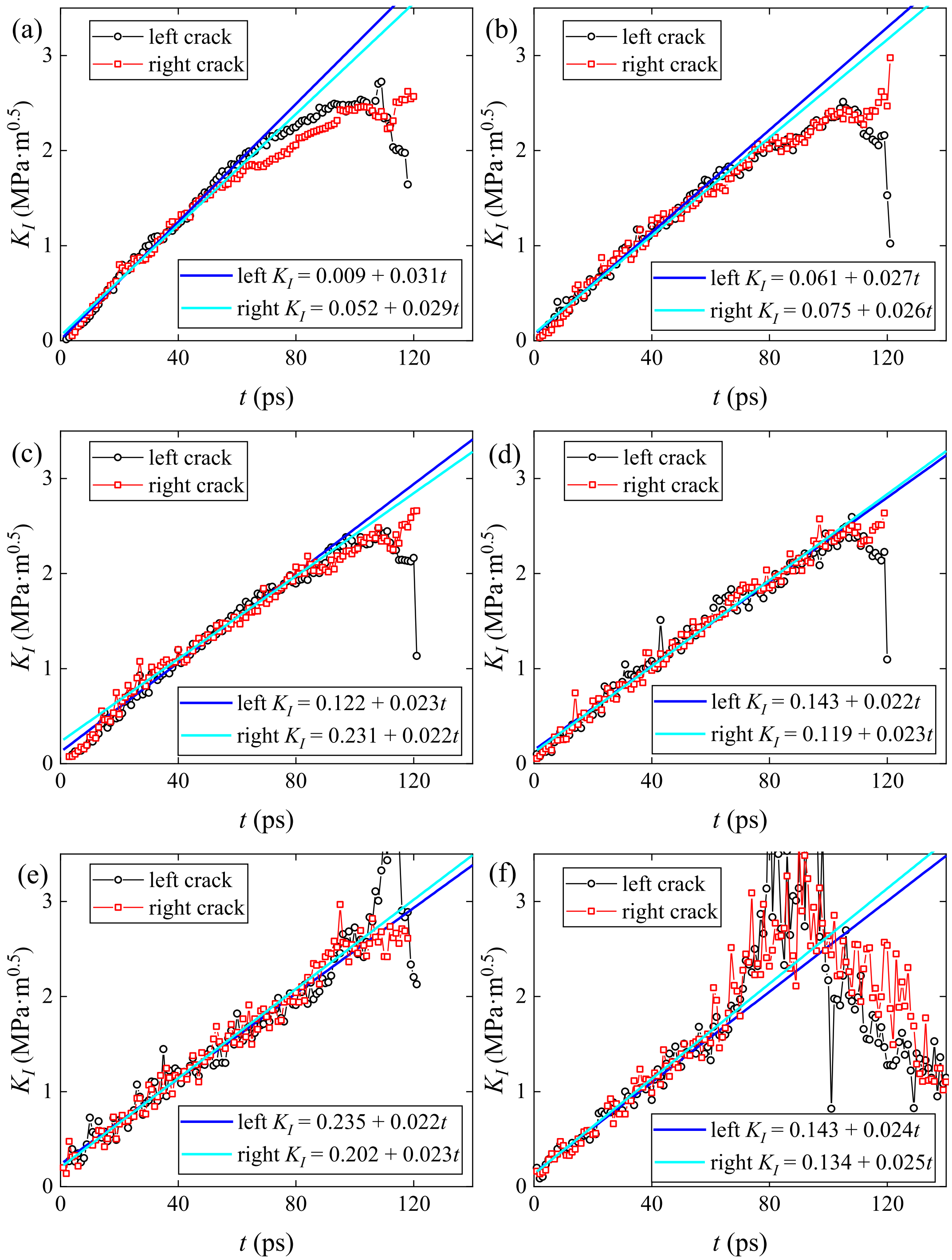

**Fig. 11**. The DSIF $K_I(t)$ as a function of time $t$ for the Σ3(109.5) specimen at (a) $T = 0.01T_m$; (b) $T = 0.1T_m$; (c) $T =$ 300 K; (d) $T = 0.3T_m$; (e) $T = 0.5T_m$; (f) $T = 0.7T_m$. For each panel, the blue and cyan curve are the linearly fitted to the left (black) and right (red) crack data points, respectively.

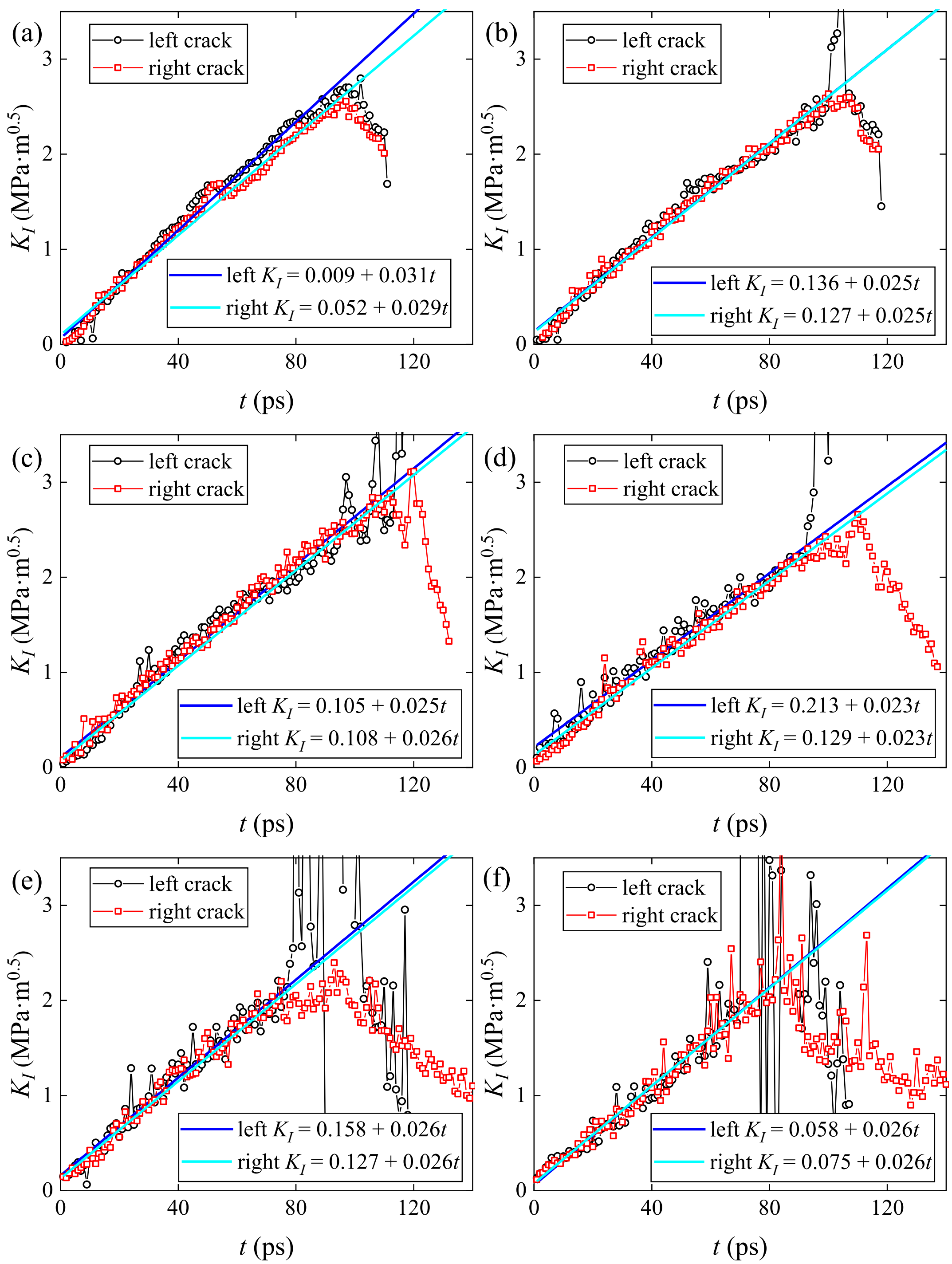

**Fig. 12**. The DSIF $K_I(t)$ as a function of time $t$ for the Σ9(141.6) specimen at (a) $T = 0.01T_m$; (b) $T = 0.1T_m$; (c) $T =$ 300 K; (d) $T = 0.3T_m$; (e) $T = 0.5T_m$; (f) $T = 0.7T_m$. For each panel, the blue and cyan curve are the linearly fitted to the left (black) and right (red) crack data points, respectively.

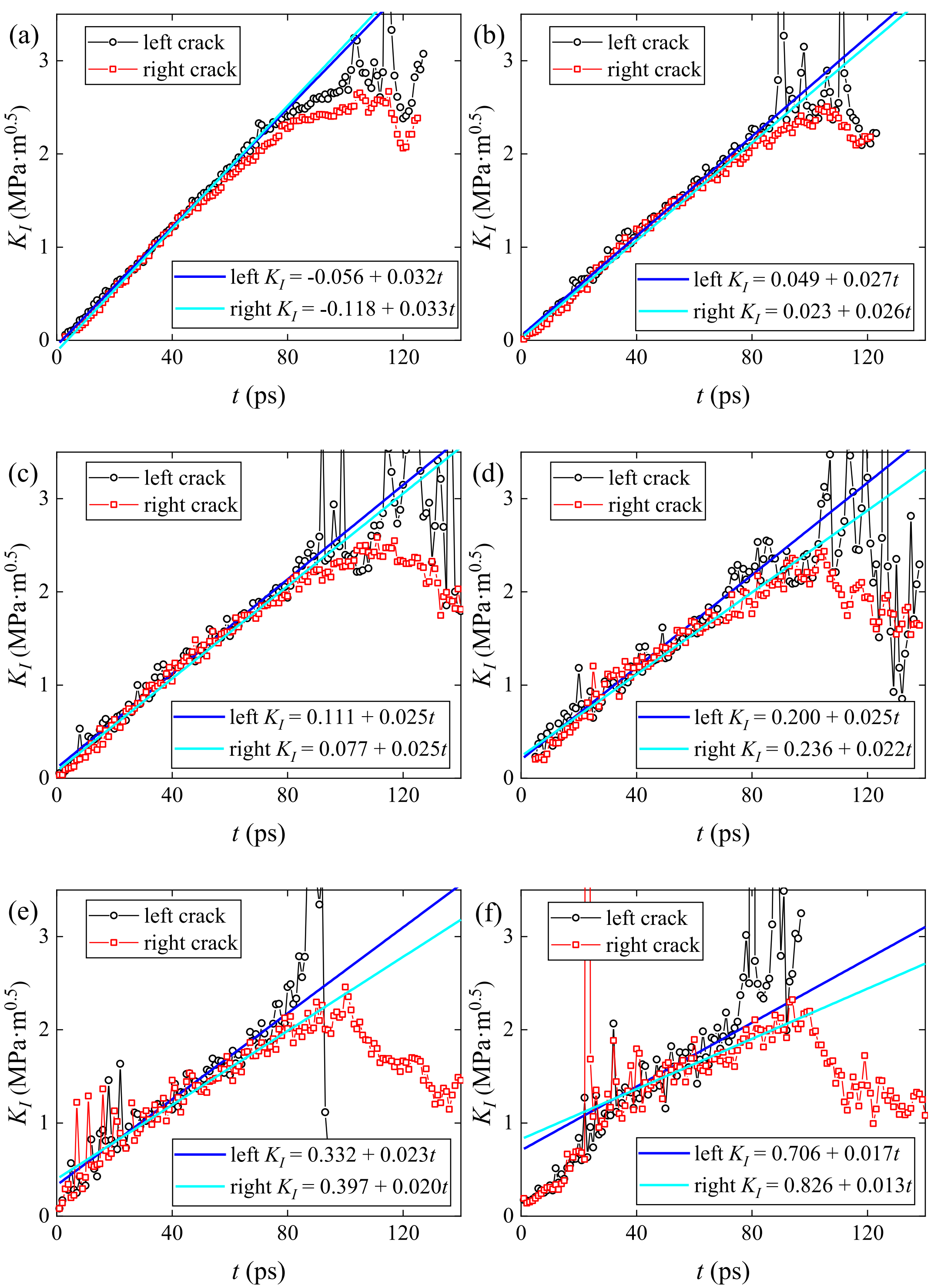

**Fig. 13**. The DSIF $K_I(t)$ as a function of time $t$ for the Σ11(129.5) specimen at (a) $T = 0.01T_m$; (b) $T = 0.1T_m$; (c) $T =$ 300 K; (d) $T = 0.3T_m$; (e) $T = 0.5T_m$; (f) $T = 0.7T_m$. For each panel, the blue and cyan curve are the linearly fitted to the left (black) and right (red) crack data points, respectively.

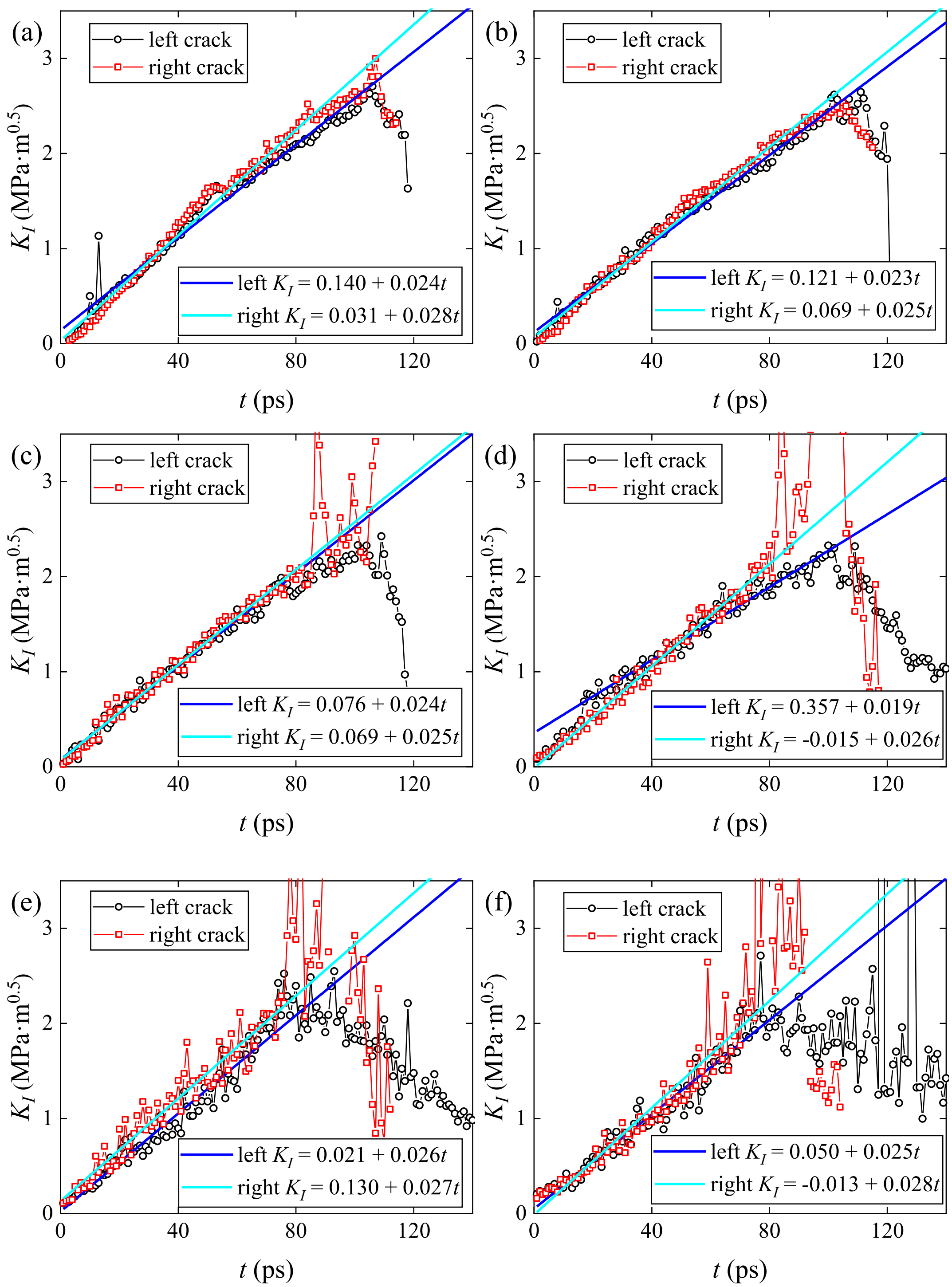

**Fig. 14**. The DSIF $K_I(t)$ as a function of time $t$ for the Σ27(148.4) specimen at (a) $T = 0.01T_m$; (b) $T = 0.1T_m$; (c) $T =$ 300 K; (d) $T = 0.3T_m$; (e) $T = 0.5T_m$; (f) $T = 0.7T_m$. For each panel, the blue and cyan curve are the linearly fitted to the left (black) and right (red) crack data points, respectively.

## 5. Discussion

In this section, the Mode-I evolution of bicrytsalline specimens is initially compared with that of counterpart single crystal specimens to illustrate the effect of GBs on the fracture behaviors. Then, the GB effect is rationalized in the theoretical framework proposed in **Section 3.2** by tuning the energy change induced by GB structural transition. Besides, the rate dependence observed in MD simulations is also discussed in details within the theoretical framework.

### *5.1 Effects of GBs on the Mode-I SIF evolution*

In order to reveal the effect of GBs on the brittle/ductile fracture behaviors of bcc Fe, we further performed the simulations of single crystal specimens with the same lattice orientations as the upper or lower grain of the bicrystalline specimens shown in **Fig. 1**a and **Table 1**. As shown in **Fig. 15**, the comparison of DSIFs between the bicrystalline and single crystal specimens indicates that depending on the GB types, the presence of GBs could retard or facilitate the crack propagation, which is marked as the deviation of the $K_I(t)$ *v.s.* $t$ curve from the linear elasticity. For instance, the sudden increase of the $K_I$ value of the single crystal specimens in **Fig. 15**a corresponds to the nucleation of twins at the crack-tip, while the fluctuation of the black curve represents the formation of amorphous region in front of the left crack-tip in the Σ3(70.00) bicrystalline specimen. It is also found that the simulated SIF rates for each specimen are in reasonable agreement with the theoretical predictions (marked by blue lines) based on anisotropic elasticity (see details in **Table B1** in **Appendix B**). The fluctuation of single crystal curves in **Fig. 15** is due to the strong dislocation activities.

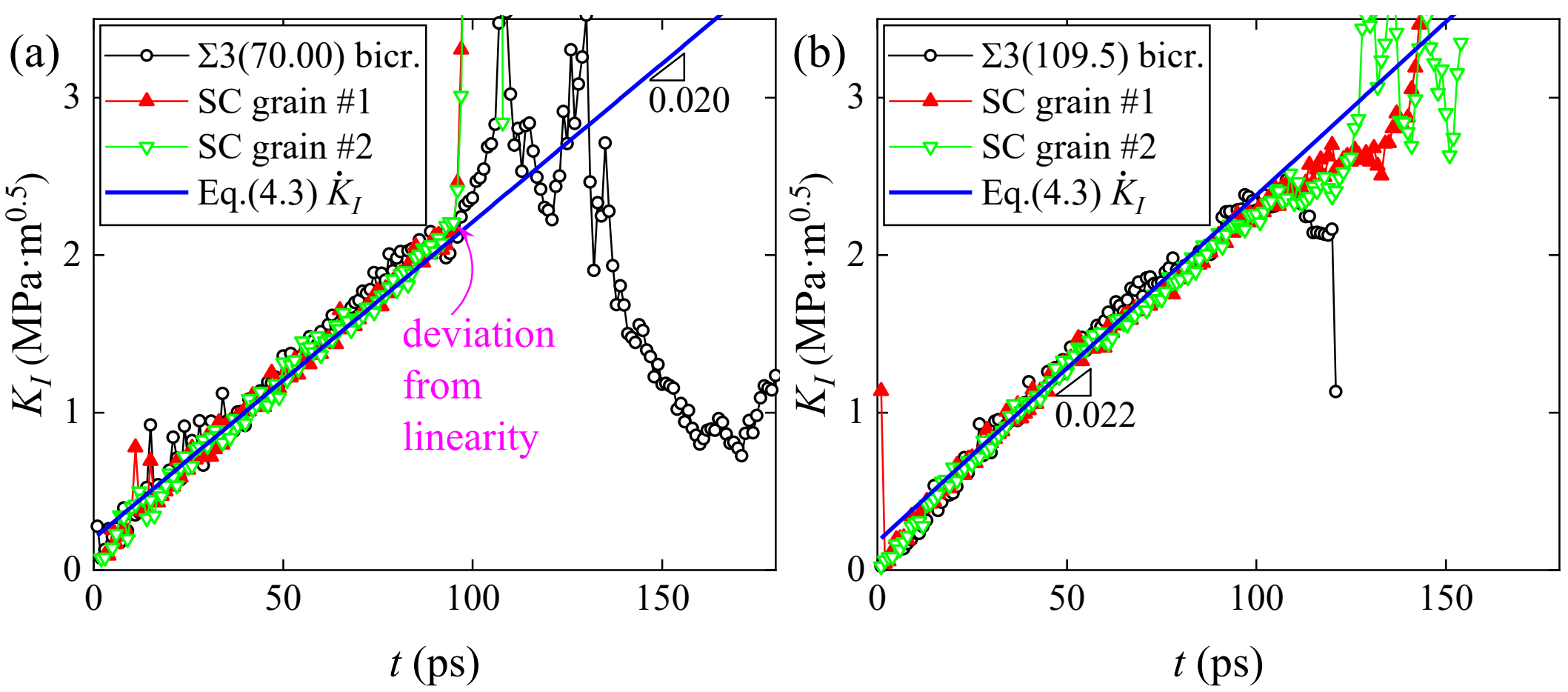

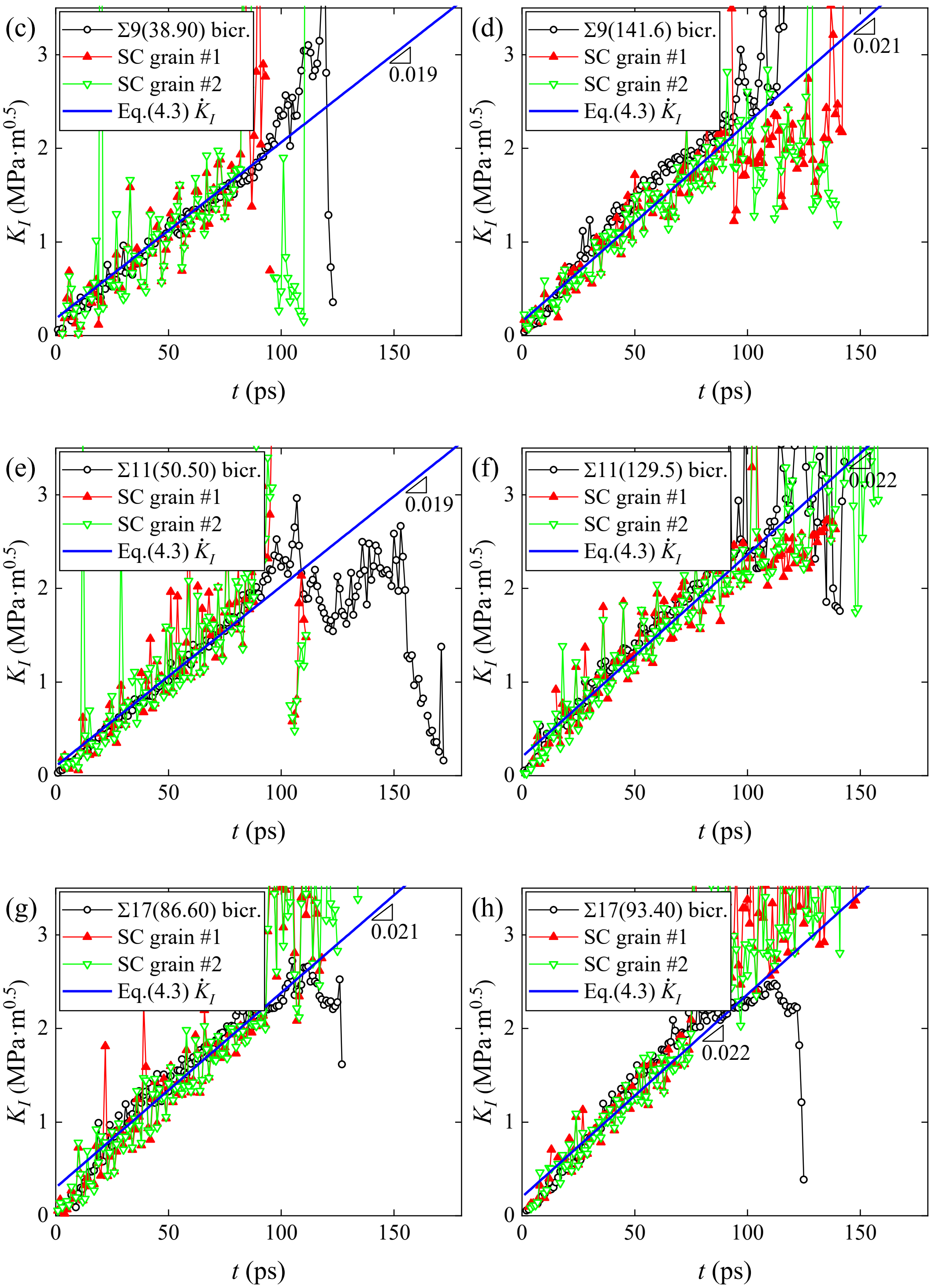

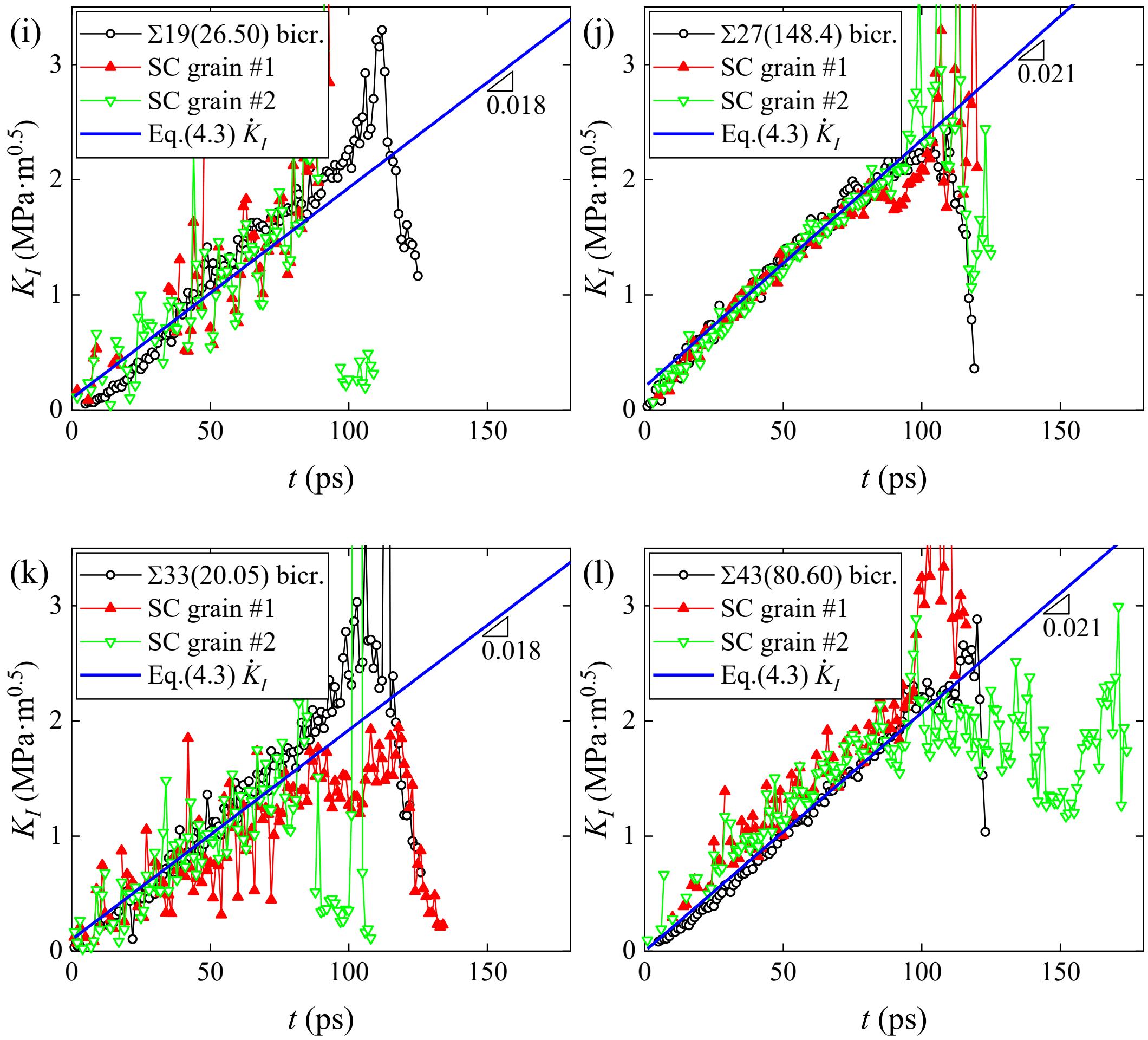

**Fig. 15**. The dynamic Mode-I SIF $K_I(t)$ of the left crack-tip as a function of the time $t$ for 12 different bicrystals and 24 corresponding single crystals with the environmental temperature $T$ = 300 K and loading rate $\dot{\varepsilon} = 5.0 \times 10^8$ /s. The blue line is theoretically evaluated by Eq.(51) based on the anisotropic elasticity (see details in **Table B1** in **Appendix B**) with the slope indicated by the triangle. It is noted that the blue line is not fitted to the MD results.

It is also found in above figures of DSIF evolution (**Fig**.s **11-14**) that the propagation behaviors of the left and right cracks differ from each other significantly. This originates from the fact that the relative distance between the crack-tip and deformation site inside the GBs is different for the left and right cracks, respectively. For the right crack in aluminum specimens, Shimokawa and Tsuboi [5] have found that the dislocation source density of the GBs increases with increasing deviation angle ($\Delta\theta$) of $\langle 112 \rangle$ tilt GBs. Thus, the plastic work ($\gamma_p$) caused by dislocation emission from GBs increases with $\Delta\theta$. In contrast, the $\gamma_p$ induced by dislocation emission from the left crack-tip decreases with increasing $\Delta\theta$. Since the GB energies adopted in their studies [5] increase with increasing $\Delta\theta$, thus the $(2\gamma_s - \gamma_{gb})$ decreases with increasing $\Delta\theta$.

These results indicate that, when GBs serve as dislocation sources, the assumption that the plastic work ($\gamma_p$) is an increasing function of ($2\gamma_s - \gamma_{gb}$) proposed by Jokl *et al*. [76] is not necessarily valid. Therefore, the role of GBs as dislocation sources cannot be ignored when analyzing the intergranular fracture behavior of nanostructured materials [77].

### *5.2 Effects of the GB structural transformation on the crack-tip plasticity*

Informed by **Fig. 15**, the deviation of the crack propagation in bicrystalline specimens with respect to single crystals might originate from the transformation of GB structural unit in the vicinity of the crack-tip [5]. As seen in **Fig. F1** in **Appendix F**, for bicrystalline specimens showing significant deviation of the SIF evolution from their counterpart single crystal specimens in **Fig. 15**, it is found that considerable transition of GB structure is observed. The key-point to consider the effect of local deformation in present theory is the introduction of the localized Mode-I SIF $K_I^*$ in Eq.(38). Before applying the theory as detailed in **Section 3**, we initially tested the classical Rice model for dislocation nucleation from a crack-tip based on anisotropic elasticity [14], which does not consider the effect of GBs and loading rate. As shown in **Table D1**, the anisotropic framework formulated by Sun and Beltz [14] underestimates the critical $K_I$ compared with the MD results, but is generally on the order of the $K_I^{max}$ evaluated from the global stress-strain curve (see **Fig. 4**), since the effect of localized atomic configuration was not considered in the classical theory. Revisiting the theoretical framework formulated in **Section 2**, it is found that the fourth term $f_{gbe}$ (specifically, the parameter $\Delta E_{gb}$) is undetermined, since the transformation of GB structure is not uniform for different specimens studied in MD simulations. In **Fig. 16**a, we parametrically calculated the normalized activation energy $\Theta_{2d}$ for various values of $\Delta E_{gb}$ (with the slip plane oriented at an angle $\theta = 0.3\pi$ to the crack plane), and found that the variation of $\Delta E_{gb}$ does not significantly change the $\Theta_{2d}$ even up to ~ $\Delta E_{gb}$ = 10 GPa·Å$^2$. It is noted that the GB transition induced $\Delta E_{gb}$ with a value of 100 GPa·Å$^2$ leads to notable elevation of $\Theta_{2d}$ compared to the case without GB transition, *i.e.* $\Delta E_{gb} = 0$. In **Fig. 16**b, we also studied the effect of the slip angle $\theta$ on the variation of the activation energy. The results show that the activation energy $\Theta_{2d}$ decreases significantly with increasing slip angle $\theta$.

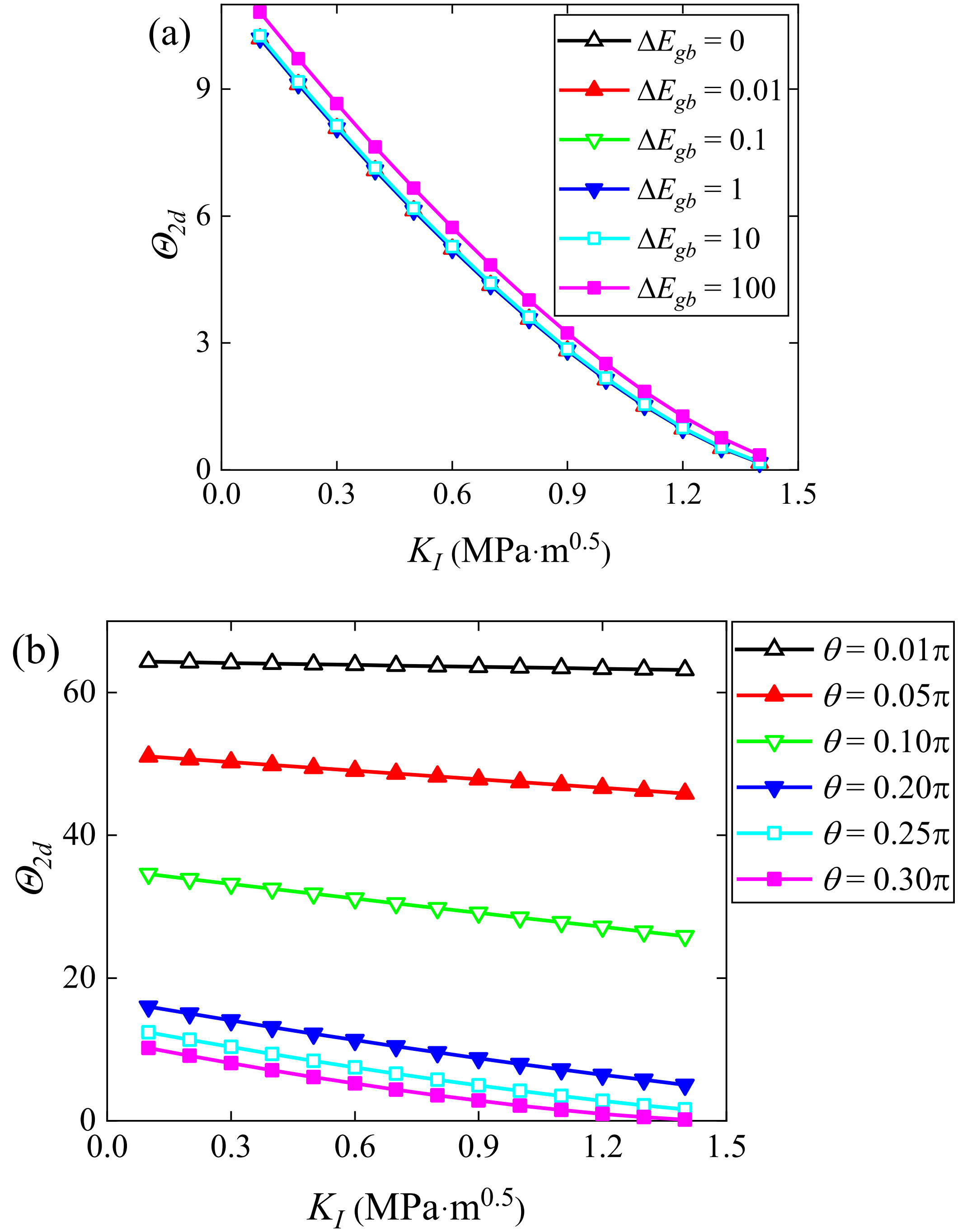

**Fig. 16**. The normalized activation energy for dislocation nucleation from the crack-tip $\Theta_{2d}$ as a function of the applied Mode-I SIF $K_I$. (a) under various GB energy change $\Delta E_{gb}$ (due to dislocation emission) values, with the angle between the slip plane and the crack plane $\theta = 0.3\pi$. The unit of the GB energy change $\Delta E_{gb}$ is GPa·Å$^2$. (b) under various $\theta$ values, with the GB energy change $\Delta E_{gb} = 0$.

Thus, without loss of generality, we attempt to solve the theoretical framework with setting the GB energy difference (between the GB1 and GB0 shown in **Fig. 2**) $\Delta\gamma_{gb}$ = 0.005 J/m$^2$, and the spacing between GB dislocation $h$ = 2 Å, *i.e.*, $\Delta E_{gb}$ = 0.1 GPa·Å$^2$. The calculated activation energy is shown in **Fig. 17**, together with the curve fitted to Eq.(45). The fitting parameters are $\tilde{C}$ = 11.31484 (±1.01384×10$^{-14}$), $K_I^0$ = 1.49036 (±3.10331×10$^{-15}$) MPa$\sqrt{\text{m}}$, and $n$ = 1.5 (±6.12064×10$^{-15}$). We also try the parameters $\Delta\gamma_{gb}$ = 0.05 J/m$^2$ and $h$ = 2 Å, *i.e.*, $\Delta E_{gb}$ = 1 GPa·Å$^2$, and obtain the fitting parameters as $\tilde{C}$ = 11.32061 (±9.26866×10$^{-15}$), $K_I^0$ = 1.49086 (±2.83812×10$^{-15}$) MPa$\sqrt{\text{m}}$, and $n$ = 1.5 (±5.59511×10$^{-15}$). It is noted that the power exponent $n$ is equal to 1.5 for both two cases, while the athermal load $K_I^0$, at which the nucleation event occurs without thermal activation, is dependent on the energy change $\Delta E_{gb}$ induced by the GB transition.

### *5.3 Rate dependence of the most probable SIF $K_{Id}^p$*

As shown in **Fig. 18**, by further solving the Eq.(47), the most probable SIF $K_{Id}^p$ can be obtained as a function of the loading rate $\dot{K}_I$. Besides, as highlighted by Guziewski *et al*. [78], it is necessary to estimate the value of the number of nucleation sites $N$ properly [71]. Since the dislocations can form anywhere within the crystal, a reasonable estimation of $N$ for homogeneous dislocation nucleation would be the number of atoms in the crystal, $n_a$. This estimation leads to a size scaling law, in which $N$ is solely the function of the number of equivalent nucleation sites, e.g., for a cubic crystal with the characteristic length $L$, $N$ would scale with $L^3$ as the cube expands self-similarly. Heterogeneous dislocation nucleation occurs at the surface (or interface) of the crystal, the number of nucleation sites can be estimated as $N = n_a A/V$ to a first approximation. In present simulations, the GB area is $A \cong L_x \times L_z$ considering the crack surface is small compared with the GB, thus $N \cong n_a/L_y$. Compared with the homogeneous dislocation nucleation, the heterogeneous nucleation would be more likely to dominate due to its lower activation energy while both the homogeneous and heterogeneous patterns coexist. Although the nucleation rate $\omega$ scales linearly with $N$, it depends exponentially on the activation energy $Q_{3d}$ and the inverse of temperature $1/T$. Thus, it is sufficient to estimate $N$ with the approximation based on the order $L_y$ and the atom number $n_a$.

However, as seen from Eq.(47), it is found that $K_{Id}^p$ is actually a function of the ratio between the loading rate and the number of the nucleation sites, *i.e.*, the rate-size parameter $\zeta \equiv \dot{K}_I/N$, which has been proposed in previous studies [79, 80] to evaluate the effect of both the spatial and temporal scale on the activation of critical events. **Fig. 18** indeed shows that the most probable SIF $K_{Id}^p$ increases monotonically with the increasing rate-size parameter $\zeta$, while the scaling factor $s_0$ is larger (*i.e.*, the 3D size effect is more significant), the sensitivity of $K_{Id}^p$ to $\zeta$ is weaker. It is also noted the $N$ corresponded size should be larger or equal to the scaling factor $s_0$. Thus, for the 3D analysis with a smaller $s_0$, the lower limit means an extremely smaller rate; while for a larger $s_0$, the upper limit means a relatively higher rate. The numerical procedure also indicates that the value of $K_{Id}^{p}$ cannot exceed the athermal load $K_I^0$, otherwise unphysical, imaginary solutions emerge. Consistent with the theoretical predictions, **Fig. 19** shows that the critical SIF corresponding to the crack-tip events (marked as the sharp transition of the DSIF curve) generally increases with the increasing loading rate. Specifically, by taking the Σ19(26.50) specimen as an example (see **Table D1** in **Appendix D**, $\theta = 48.5°$), the theoretical predictions (**Fig. 18**b) agree well with the MD results (**Fig. 19**i) while the energy change $\Delta E_{gb} \approx 600$ GPa·Å$^2$, which means a dramatic alternation of GB structure was anticipated.

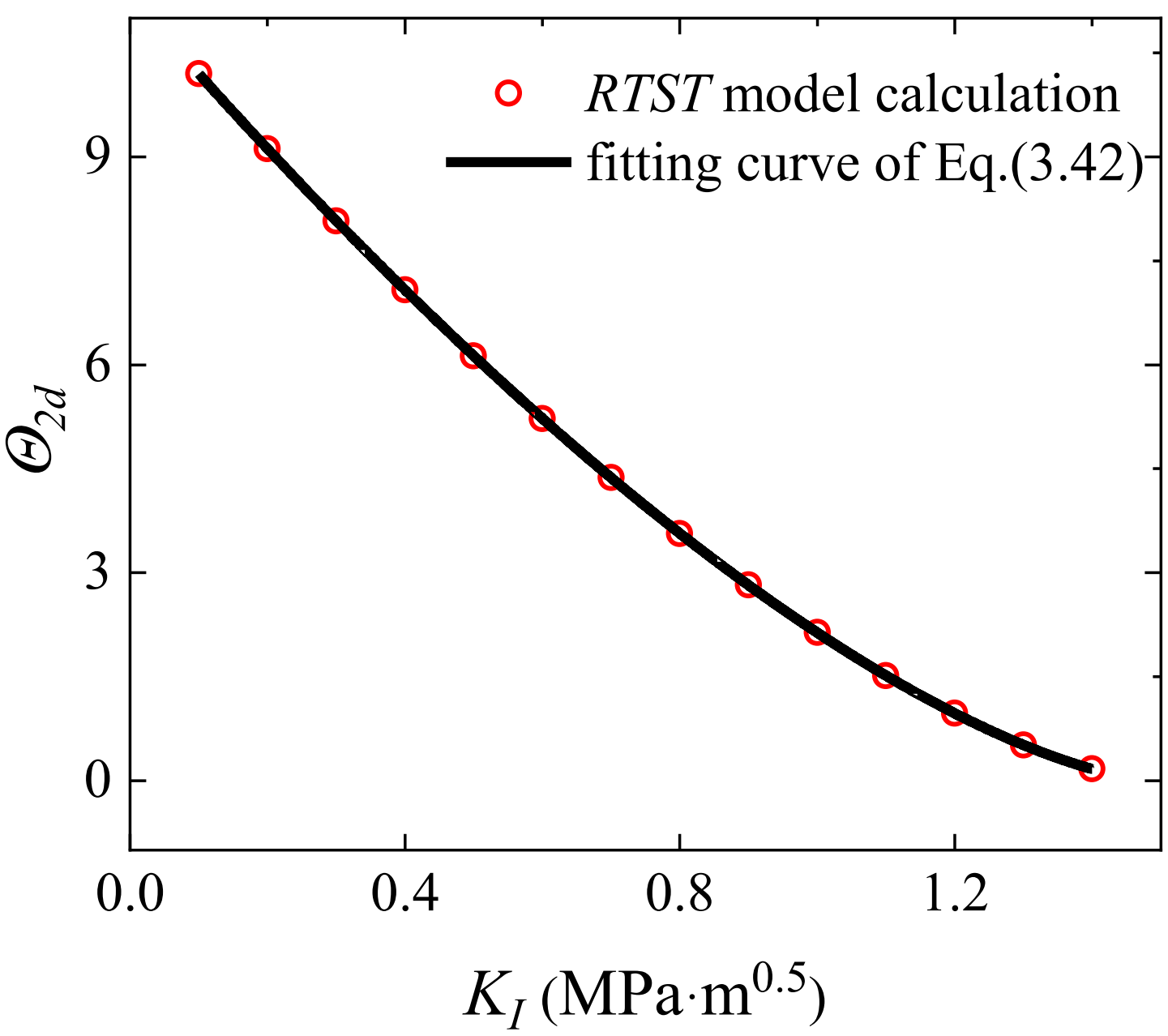

**Fig. 17**. The schematics of the $\Theta_{2d}(K_I)$ *v.s.* $K_I$ fitting procedure. The red circle represents the theoretical predictions via the ***RTST*** and ***PRB*** models, while the black curve is obtained by fitting the theoretical results to Eq.(45).

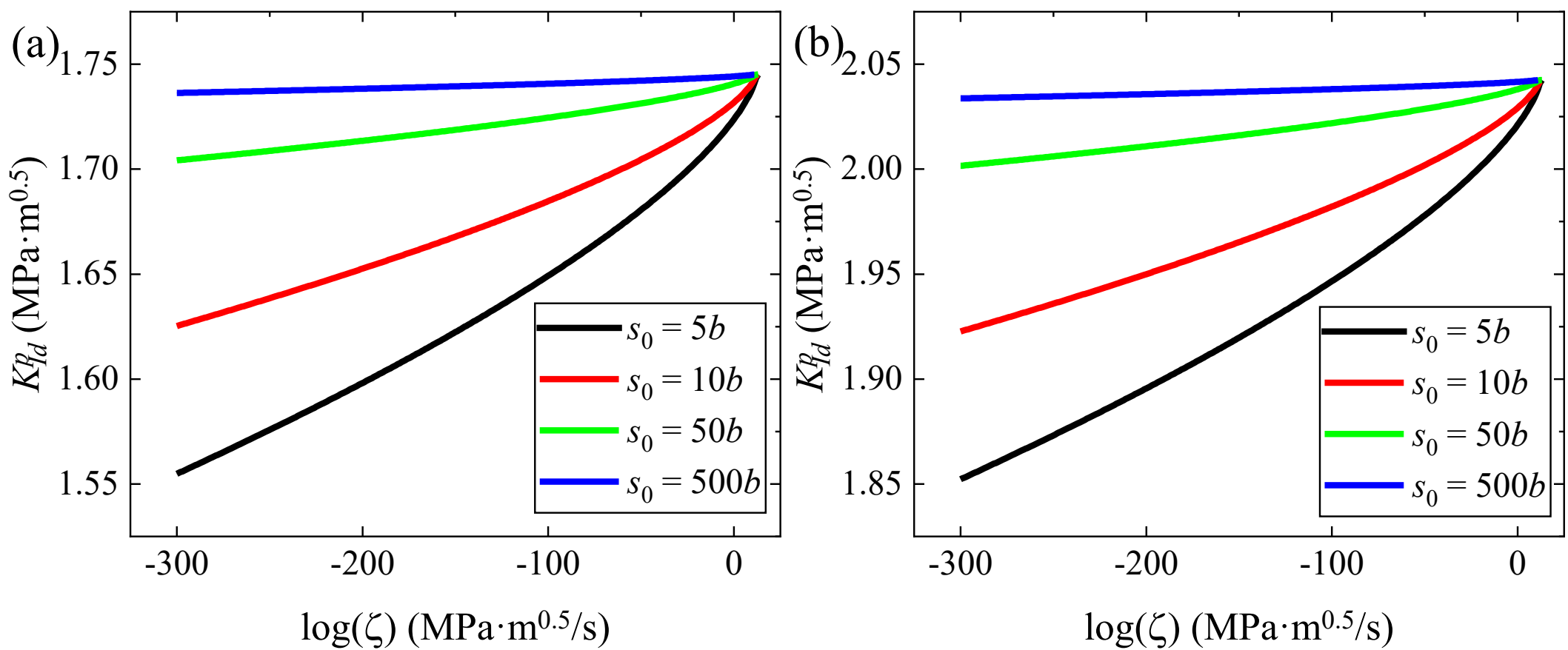

**Fig. 18**. The $K_{Id}^{p}$ *v.s.* $\zeta$ relation with various scaling factor $s_0$ with the angle $\theta$ = 48.5° and, (a) $\Delta E_{gb}$ = 100 GPa·Å$^2$, (b) $\Delta E_{gb}$ = 600 GPa·Å$^2$.

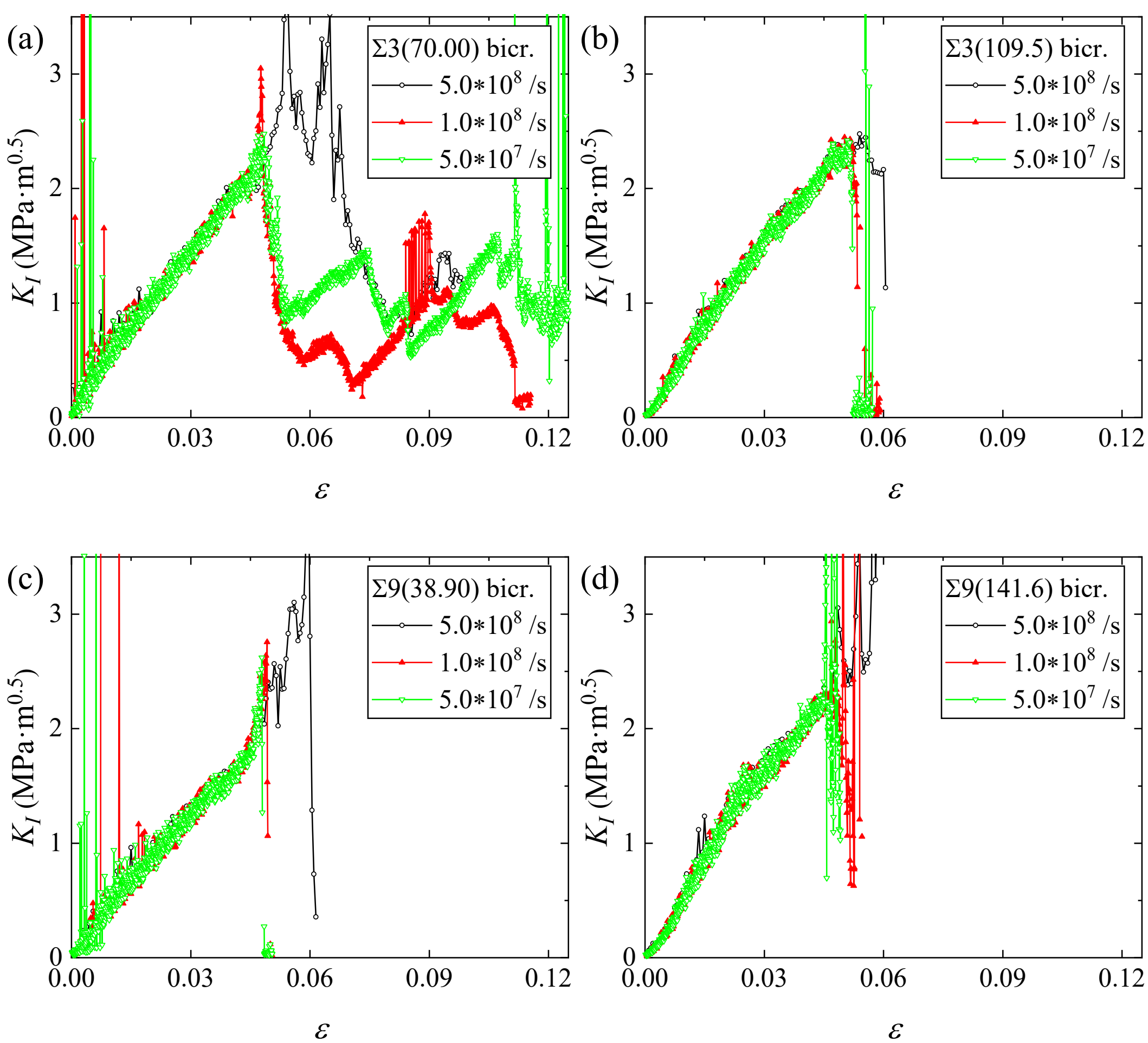

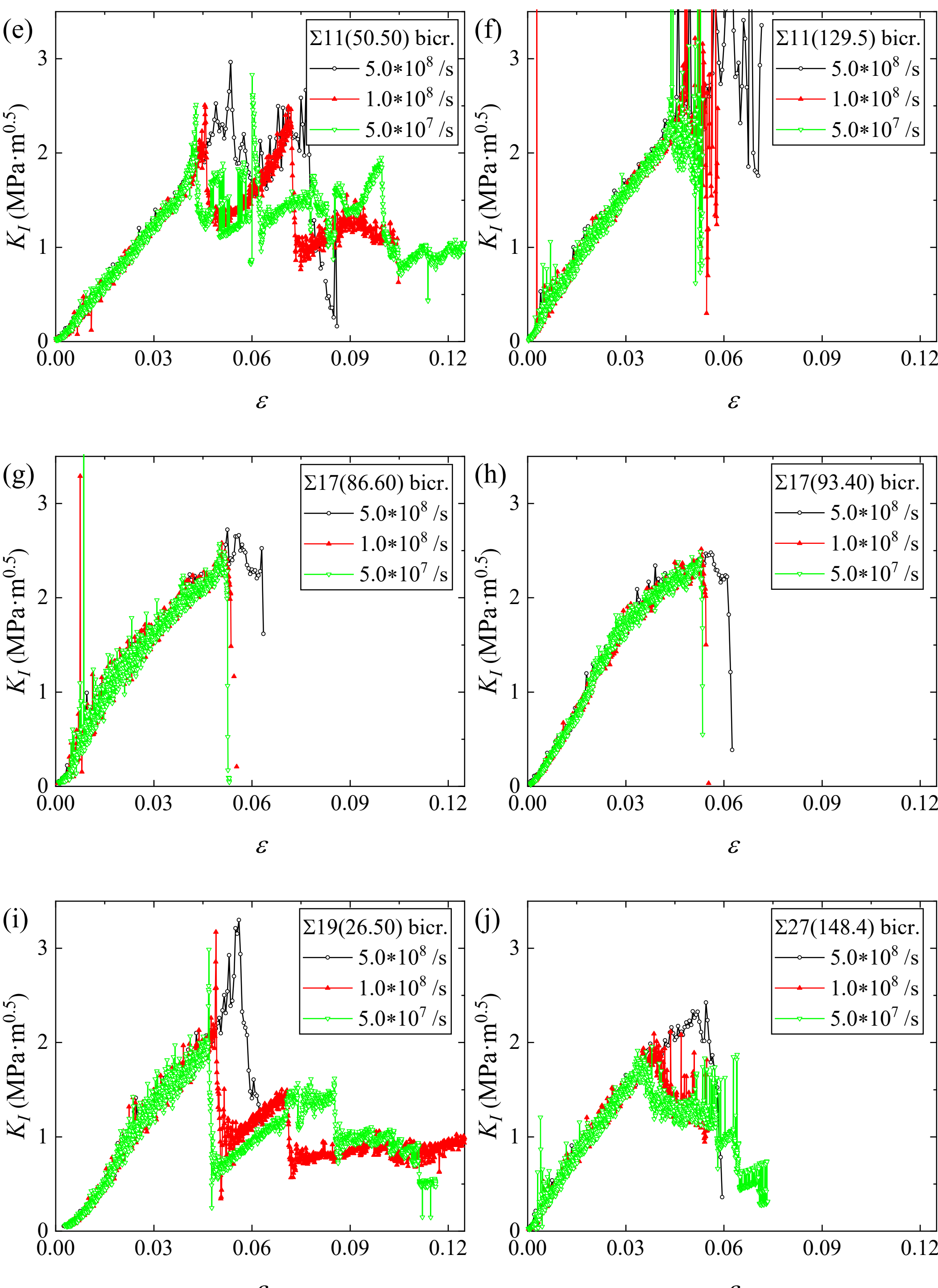

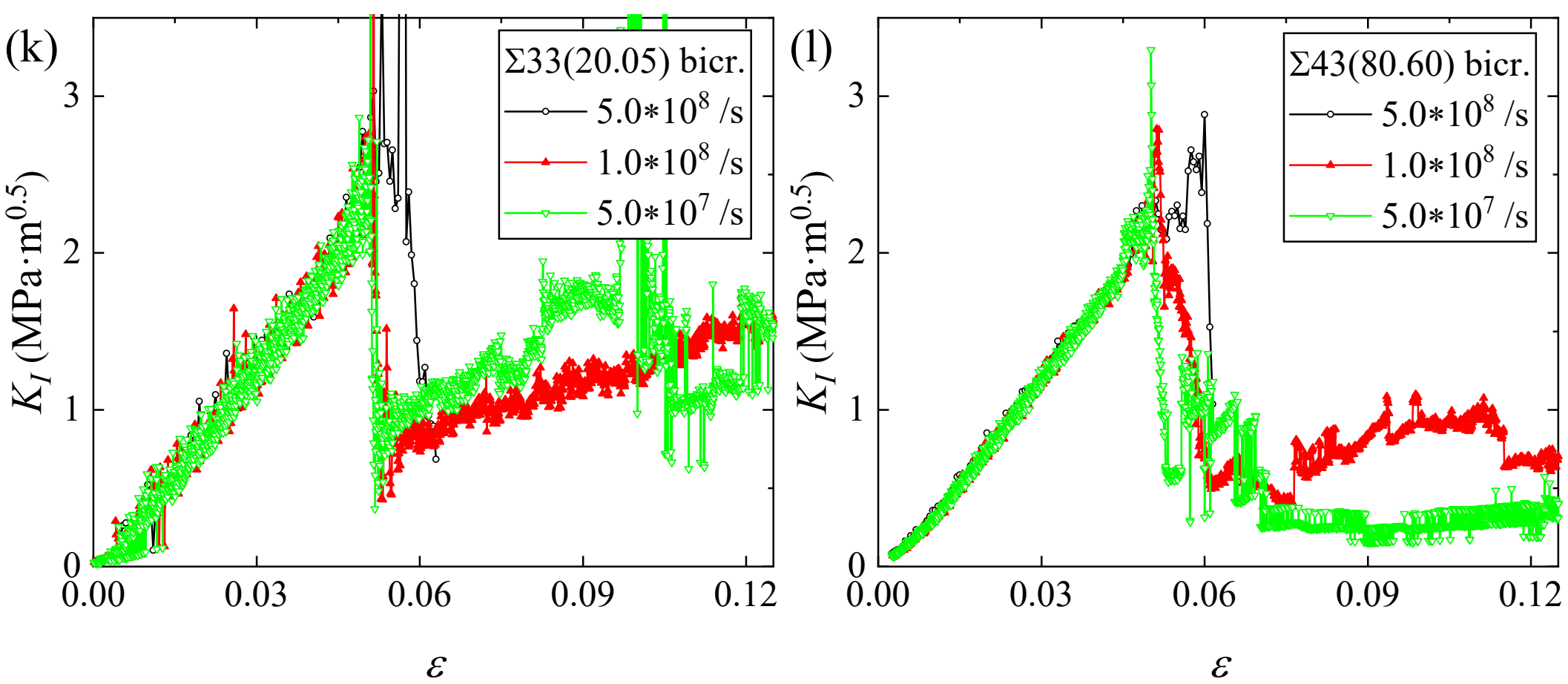

**Fig. 19**. The dynamic Mode-I SIF $K_I$ of the left crack-tip as a function of the strain $\varepsilon$ for 12 different bicrystals under different loading rates ($\dot{\varepsilon} = 5.0 \times 10^8, 1.0 \times 10^8, 5.0 \times 10^7$ /s, respectively) with the environmental temperature $T$ = 300 K.

As suggested in Ref.s[13, 35], we use a scaling factor $s_0 = 5b$ (with the angle $\theta$ = 0.3π) to study the influence of GB dislocation emission on $K_{Id}^p$. As shown in **Fig. 20**, the transition of GB structures has negligible influence on the most probable SIF $K_{Id}^p$ when $|\Delta E_{gb}|$ is smaller than ~ 1.0 GPa·Å$^2$. However, $K_{Id}^p$ changed significantly when $|\Delta E_{gb}|$ is larger than ~ 10 GPa·Å$^2$. For the GBs studied here (see **Table C1**), the maximum of $|\Delta \gamma_{gb}|$ could be 1.308 J/m$^2$ referred to single crystal, if we consider the spacing between GB dislocations $h$ = 10 Å, $|\Delta E_{gb}|$ can be evaluated as 130.8 GPa·Å$^2$. Therefore, the contribution due to GB structural transition cannot be neglected when studying the incipient plasticity under the ***PRB*** framework. It is noted that the GB transition with a positive $\Delta E_{gb}$ would lead to the elevation of $K_{Id}^p$ (see **Fig. 20**a). By the same token, **Fig. 20**b shows that a negative $\Delta E_{gb}$ leads to the decrease of $K_{Id}^p$. However, there seems to exist considerable differences between the predicted $K_{Id}^p$ values and MD results shown in **Fig. 15**, mainly due to the anisotropy of the specimen configuration, *i.e.*, the variation of the slip angle $\theta$ between the slip plane and the crack plane. As shown in **Fig. 21**, $K_{Id}^p$ changes significantly but not monotonically with the angle $\theta$ (with $\Delta E_{gb} = 0$). Besides, the surface energy $\gamma_{surf}$ [81, 82] and USF energy $\gamma_{usf}$ [83, 84] used in the theoretical calculations change upon the variation of the lattice orientation. Previous studies [85] also revealed that the generalized stacking fault energy (GSFE) curve is affected by the resolved stress normal to the slip plane. Thus, during the dynamic loading process, the values of $\gamma_{usf}$ and $\gamma_{surf}$ should be modified simultaneously for different loading states and orientations.

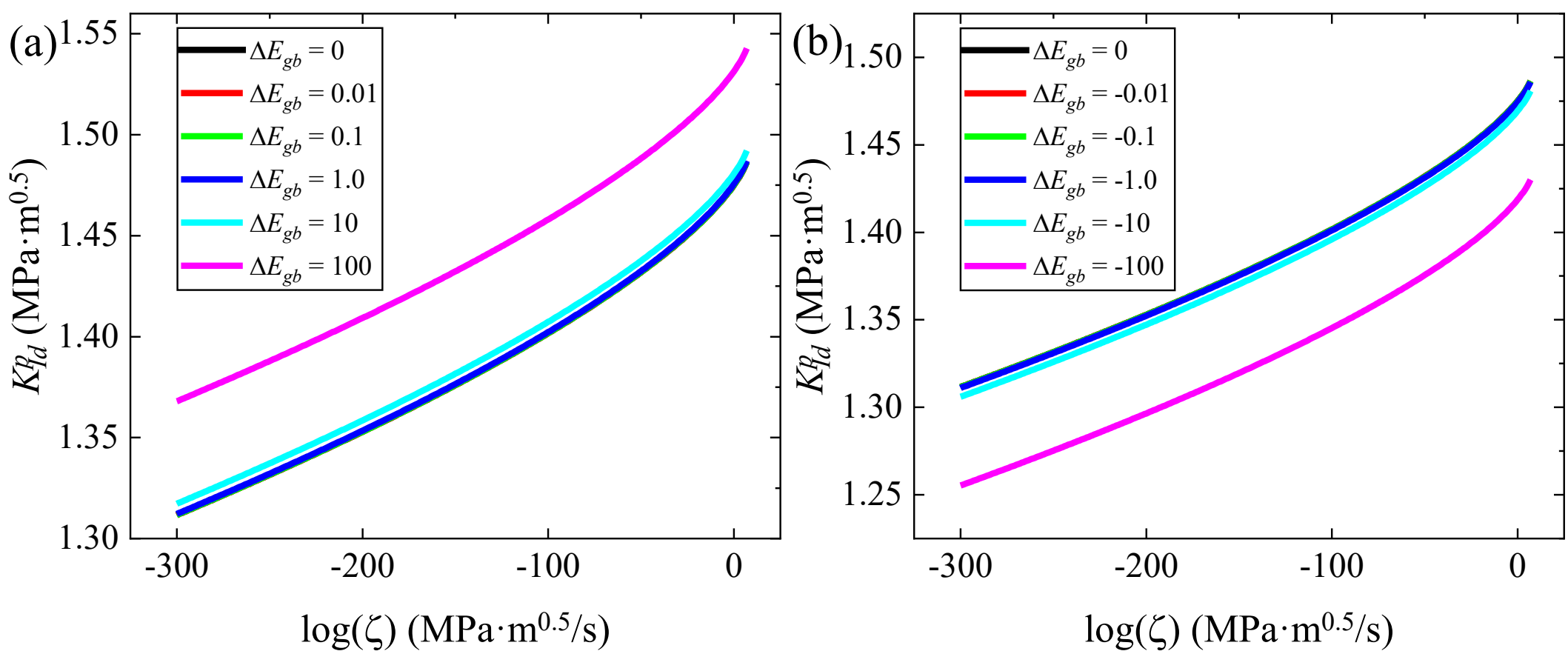

**Fig. 20**. The $K_{Id}^{p}$ *v.s.* $\zeta$ relation with the scaling factor $s_0 = 5b$ (with the angle $\theta = 0.3\pi$) for cases with different $\Delta E_{gb}$ (in the unit of GPa·Å$^2$). (a) for $\Delta E_{gb} > 0$; (b) for $\Delta E_{gb} < 0$.

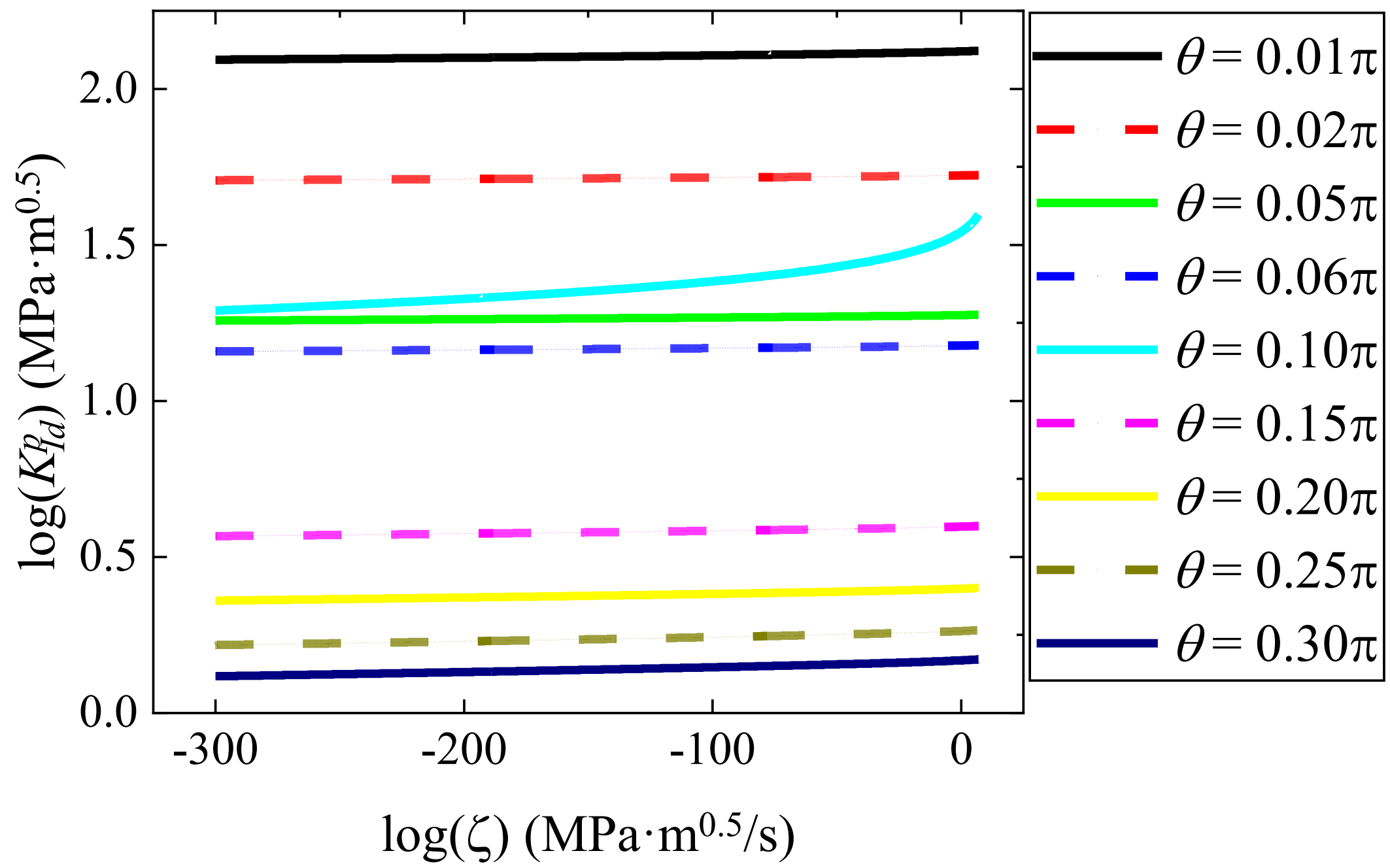

**Fig. 21**. The $K_{Id}^{p}$ v.s. $\zeta$ relation with the scaling factor $s_0 = 5b$ (with $\Delta E_{gb} = 0$) for cases with different angles $\theta$.

# 6. Conclusions

By extending the classical Rice model to consider the GB transition, we established a theoretical framework to predict the crack-tip dislocation emission during the intergranular fracture process, and verified it by using large-scale atomistic simulations for bicrystalline specimens with various GB characters. The following conclusions can be drawn as, 1) Compared to both the $K_I^{max}$ evaluated from the global stress-strain curve and the $K_I^e$ predicted by the classical Rice model, the measurement of the dynamic stress intensity factor (SIF) indicates that the $K_I$ field was enhanced near the crack-tip, thus the localized variation of the atomic configuration has to be considered. 2) With respect to the tests of single crystal specimens, the existence of GB in bicrystalline specimens does change the critical SIF for dislocation emission due to the transformation of GB structural unit. In other words, the deviation from Rice model originates from the competition of dislocation nucleated from either the crack-tip or GB sources. 3) While classical models underestimate the MD results under high loading rate, the present theoretical framework successfully predicts the increase in critical SIF with the rising loading rate.

Although the present study was conducted for the bcc metal, it is anticipated that the theoretical framework also applies to fcc metals, while the difference in the GSFE curves between bcc and fcc lattices should be considered. By further considering other deformation patterns, such as twining, phase transition, nanovoiding and amorphization, the present framework sheds a light to the mechanism-based design and manufacturing of GB-engineered materials with optimized strength and ductility.

## Data availability statement

The raw data required to reproduce these findings are available upon a reasonable request. Please directly contact the corresponding author of this paper.

## CRediT authorship contribution statement

**K. Zhao**: Conceptualization, Writing – original draft, Writing – review & editing, Project administration, Funding acquisition. **Y. Ding**: Writing – review & editing. **H. Yu**: Writing – review & editing, **J. He**: Writing – review & editing, Project administration, Funding acquisition. **Z. Zhang**: Writing – original draft, Writing – review & editing, Project administration, Funding acquisition.

## Declaration of competing interest

The authors declare that they have no known competing financial interests or personal relationships that could have appeared to influence the work reported in this paper.

## Acknowledgements

Financial supports provided by the National Natural Science Foundation of China (Grant No. 12102145), Natural Science Foundation of Jiangsu Province (Grant No. BK20210444), Entrepreneurship and Innovation Doctor Program of Jiangsu Province (Grant No. JSSCBS20210856), the Fundamental Research Funds for the Central Universities (Grant No. JUSRP121042), and the Research Council of Norway (Grant No. 347726, 344377, 344297) are acknowledged.

## Appendix A. Determination of the anisotropic elastic coefficients

Established by Lieberman and Zirinsky [56, 86], the elastic constants of arbitrary oriented single crystals can be determined by the following procedures,

1) determine the direction cosines connecting the two unitary orthogonal axis systems as:

$$\begin{bmatrix} x_1' \\ x_2' \\ x_3' \end{bmatrix} = \begin{bmatrix} \beta_{11} & \beta_{12} & \beta_{13} \\ \beta_{21} & \beta_{22} & \beta_{23} \\ \beta_{31} & \beta_{32} & \beta_{33} \end{bmatrix} \begin{bmatrix} x_1 \\ x_2 \\ x_3 \end{bmatrix} \tag{A.1}$$

2) write the quadratic combinations of the direction cosines in the following 6×6 matrix:

$$\gamma = \begin{bmatrix} \beta_{11}^2 & \beta_{12}^2 & \beta_{13}^2 & \beta_{12}\beta_{13} & \beta_{13}\beta_{11} & \beta_{11}\beta_{12} \\ \beta_{21}^2 & \beta_{22}^2 & \beta_{23}^2 & \beta_{22}\beta_{23} & \beta_{23}\beta_{21} & \beta_{21}\beta_{22} \\ \beta_{31}^2 & \beta_{32}^2 & \beta_{33}^2 & \beta_{32}\beta_{33} & \beta_{33}\beta_{31} & \beta_{31}\beta_{32} \\ 2\beta_{21}\beta_{31} & 2\beta_{22}\beta_{32} & 2\beta_{23}\beta_{33} & \begin{pmatrix} \beta_{22}\beta_{33} + \\ \beta_{23}\beta_{32} \end{pmatrix} & \begin{pmatrix} \beta_{21}\beta_{33} + \\ \beta_{23}\beta_{31} \end{pmatrix} & \begin{pmatrix} \beta_{22}\beta_{31} + \\ \beta_{21}\beta_{32} \end{pmatrix} \\ 2\beta_{31}\beta_{11} & 2\beta_{32}\beta_{12} & 2\beta_{33}\beta_{13} & \begin{pmatrix} \beta_{13}\beta_{32} + \\ \beta_{12}\beta_{33} \end{pmatrix} & \begin{pmatrix} \beta_{13}\beta_{31} + \\ \beta_{11}\beta_{33} \end{pmatrix} & \begin{pmatrix} \beta_{11}\beta_{32} + \\ \beta_{12}\beta_{31} \end{pmatrix} \\ 2\beta_{11}\beta_{21} & 2\beta_{12}\beta_{22} & 2\beta_{13}\beta_{23} & \begin{pmatrix} \beta_{12}\beta_{23} + \\ \beta_{13}\beta_{22} \end{pmatrix} & \begin{pmatrix} \beta_{13}\beta_{21} + \\ \beta_{11}\beta_{23} \end{pmatrix} & \begin{pmatrix} \beta_{11}\beta_{22} + \\ \beta_{12}\beta_{21} \end{pmatrix} \end{bmatrix} \tag{A.2}$$

for the transformation of the elastic compliance matrix $S_{ij}$, and

$$\alpha = \begin{bmatrix} \beta_{11}^2 & \beta_{12}^2 & \beta_{13}^2 & 2\beta_{12}\beta_{13} & 2\beta_{13}\beta_{11} & 2\beta_{11}\beta_{12} \\ \beta_{21}^2 & \beta_{22}^2 & \beta_{23}^2 & 2\beta_{22}\beta_{23} & 2\beta_{23}\beta_{21} & 2\beta_{21}\beta_{22} \\ \beta_{31}^2 & \beta_{32}^2 & \beta_{33}^2 & 2\beta_{32}\beta_{33} & 2\beta_{33}\beta_{31} & 2\beta_{31}\beta_{32} \\ \beta_{21}\beta_{31} & \beta_{22}\beta_{32} & \beta_{23}\beta_{33} & \begin{pmatrix}\beta_{22}\beta_{33} + \\ \beta_{23}\beta_{32}\end{pmatrix} & \begin{pmatrix}\beta_{21}\beta_{33} + \\ \beta_{23}\beta_{31}\end{pmatrix} & \begin{pmatrix}\beta_{22}\beta_{31} + \\ \beta_{21}\beta_{32}\end{pmatrix} \\ \beta_{31}\beta_{11} & \beta_{32}\beta_{12} & \beta_{33}\beta_{13} & \begin{pmatrix}\beta_{13}\beta_{32} + \\ \beta_{12}\beta_{33}\end{pmatrix} & \begin{pmatrix}\beta_{13}\beta_{31} + \\ \beta_{11}\beta_{33}\end{pmatrix} & \begin{pmatrix}\beta_{11}\beta_{32} + \\ \beta_{12}\beta_{31}\end{pmatrix} \\ \beta_{11}\beta_{21} & \beta_{12}\beta_{22} & \beta_{13}\beta_{23} & \begin{pmatrix}\beta_{12}\beta_{23} + \\ \beta_{13}\beta_{22}\end{pmatrix} & \begin{pmatrix}\beta_{13}\beta_{21} + \\ \beta_{11}\beta_{23}\end{pmatrix} & \begin{pmatrix}\beta_{11}\beta_{22} + \\ \beta_{12}\beta_{21}\end{pmatrix} \end{bmatrix} \quad \text{(A.3)}$$

for the transformation of the elastic constant matrix $C_{ij}$. The elastic coefficients of bcc Fe in the basic coordinate system ($x$-[100], $y$-[010], $z$-[001]) are: $C_{11}$ = 243.36 GPa; $C_{12}$ = 145.01 GPa; $C_{44}$ = 116.04 GPa [87].

3) calculate the transformed matrix:

$$S' = \gamma S \gamma^T \quad \text{(A.4)}$$

$$C' = \alpha C \alpha^T \quad \text{(A.5)}$$

where, $\gamma^T$ and $\alpha^T$ are the transpose matrix of $\gamma$ and $\alpha$, respectively. The transformed compliance coefficients $S'_{ij}$ for 12 different orientations are listed in **Table A1**.

**Table A1**. Compliance coefficients $S'_{ij}$ of interest for different orientations ($10^{-3}$/GPa).

| orientations | $S'_{11}$ | $S'_{12}$ | $S'_{13}$ | $S'_{22}$ | $S'_{23}$ | $S'_{33}$ | $S'_{63}$ | $S'_{66}$ |
|---|---|---|---|---|---|---|---|---|
| (112)/[$1\bar{1}0$] | 3.4975 | -0.8113 | -0.8113 | 4.4740 | -1.7878 | 4.4740 | 2.7619 | 16.4296 |
| (111)/[$1\bar{1}0$] | 4.4740 | -0.8113 | -1.7878 | 3.4975 | -0.8113 | 4.4740 | 2.7619 | 16.4296 |
| (114)/[$1\bar{1}0$] | 3.9315 | -1.8963 | -0.1604 | 6.2100 | -2.4388 | 4.4740 | 1.8413 | 12.0897 |
| (221)/[$1\bar{1}0$] | 6.2100 | -1.8963 | -2.4388 | 3.9315 | -0.1604 | 4.4740 | 1.8413 | 12.0897 |
| (113)/[$1\bar{1}0$] | 3.6993 | -1.4570 | -0.3675 | 5.5635 | -2.2317 | 4.4740 | 2.2597 | 13.8472 |
| (332)/[$1\bar{1}0$] | 5.5635 | -1.4570 | -2.2317 | 3.6993 | -0.3675 | 4.4740 | 2.2597 | 13.8472 |
| (223)/[$1\bar{1}0$] | 3.6631 | -0.5748 | -1.2134 | 3.8354 | -1.3857 | 4.4740 | 2.9244 | 17.3757 |
| (334)/[$1\bar{1}0$] | 3.8354 | -0.5748 | -1.3857 | 3.6631 | -1.2134 | 4.4740 | 2.9244 | 17.3757 |
| (116)/[$1\bar{1}0$] | 4.1900 | -2.3261 | 0.0110 | 6.8111 | -2.6101 | 4.4740 | 1.3083 | 10.3705 |
| (552)/[$1\bar{1}0$] | 6.5837 | -2.1615 | -2.5473 | 4.0882 | -0.0519 | 4.4740 | 1.5344 | 11.0288 |
| (118)/[$1\bar{1}0$] | 4.3045 | -2.5061 | 0.0764 | 7.0564 | -2.6755 | 4.4740 | 1.0043 | 9.6507 |
| (335)/[$1\bar{1}0$] | 3.5614 | -0.6254 | -1.0611 | 4.0383 | -1.5380 | 4.4740 | 2.8904 | 17.1732 |

## Appendix B. classical anisotropic linear elastic crack-tip fields

As demonstrated by Zhang *et al*. [75], the displacement and stress fields of a half crack in an infinite anisotropic medium can be derived by the ***Lekhnitskii formalism***. Detailed descriptions and the application to cracks are given by Sih *et al*. [55]. Here we briefly review the main formulas, where the governing equation of the plane strain problem in an anisotropic linear elastic medium is,

$$b_{11}\frac{\partial^4 U}{\partial y^4}+b_{22}\frac{\partial^4 U}{\partial x^4}+(2b_{12}+b_{66})\frac{\partial^4 U}{\partial x^2\partial y^2}-2b_{16}\frac{\partial^4 U}{\partial x\partial y^3}-2b_{26}\frac{\partial^4 U}{\partial x^3\partial y}=0 \tag{B.1}$$

where $U$ is the ***Airy stress function*** and $b_{ij}$ are correlated with the orientation-dependent compliance moduli $S'_{mn}$ determined by the Eq.(A.4),

$$b_{ij}=S'_{ij}-\frac{S'_{i3}S'_{j3}}{S'_{33}} \tag{B.2}$$

The calculated $b_{ij}$ coefficients for 12 different orientations are listed in **Table B1**. The characteristic equation of the governing equation Eq.(B.1) is,

$$b_{11}\mu_j^4-2b_{16}\mu_j^3+(2b_{12}+b_{66})\mu_j^2-2b_{26}\mu_j+b_{22}=0 \tag{B.3}$$

By denoting the roots $\mu_i$ ($i=1\sim4$, $\mu_1=\bar{\mu}_3$ and $\mu_2=\bar{\mu}_4$) and introducing two variables $s_1=\mu_1$, $s_2=\mu_2$, the ***Airy stress function*** for general plane strain problem can expressed as,

$$U(x,y)=2\Re[U_1(z_1)+U_2(z_2)] \tag{B.4}$$

where $z_i=x+s_iy$ ($i$ = 1, 2), $U_1$ and $U_2$ are arbitrary functions to be determined by the boundary conditions. Hence, the displacement fields can be expressed as,

$$\begin{aligned}u_x&=2\Re[p_1\phi(z_1)+p_2\psi(z_2)],\\ u_y&=2\Re[q_1\phi(z_1)+q_2\psi(z_2)].\end{aligned} \tag{B.5}$$

where $p_i$ and $q_i$ are of the form,

$$\begin{aligned}p_1&=b_{11}s_1^2+b_{12}-b_{16}s_1,\quad p_2=b_{11}s_2^2+b_{12}-b_{16}s_2,\\ q_1&=\frac{b_{12}s_1^2+b_{22}-b_{26}s_1}{s_1},\qquad q_2=\frac{b_{12}s_2^2+b_{22}-b_{26}s_2}{s_2}.\end{aligned} \tag{B.6}$$

By considering the boundary condition of a crack in an infinite medium, the stress function $\phi$ and $\psi$ can be determined. Thus, the displacement fields at the crack-tip, expressed in polar coordinates ($\theta$, $r$), can be written as,

$$u_x = K_I\sqrt{\frac{2r}{\pi}}\Re\left[\frac{1}{s_1-s_2}\left(s_1p_2\sqrt{\cos\theta+s_2\sin\theta}-s_2p_1\sqrt{\cos\theta+s_1\sin\theta}\right)\right],$$
$$u_y = K_I\sqrt{\frac{2r}{\pi}}\Re\left[\frac{1}{s_1-s_2}\left(s_1q_2\sqrt{\cos\theta+s_2\sin\theta}-s_2q_1\sqrt{\cos\theta+s_1\sin\theta}\right)\right]. \quad (B.7)$$

and the stress fields are,

$$\sigma_{xx} = \frac{K_I}{\sqrt{2\pi r}}\Re\left[\frac{s_1s_2}{s_1-s_2}\left(\frac{s_2}{\sqrt{\cos\theta+s_2\sin\theta}}-\frac{s_1}{\sqrt{\cos\theta+s_1\sin\theta}}\right)\right],$$
$$\sigma_{yy} = \frac{K_I}{\sqrt{2\pi r}}\Re\left[\frac{1}{s_1-s_2}\left(\frac{s_1}{\sqrt{\cos\theta+s_2\sin\theta}}-\frac{s_2}{\sqrt{\cos\theta+s_1\sin\theta}}\right)\right], \quad (B.8)$$
$$\sigma_{xy} = \frac{K_I}{\sqrt{2\pi r}}\Re\left[\frac{s_1s_2}{s_1-s_2}\left(\frac{1}{\sqrt{\cos\theta+s_1\sin\theta}}-\frac{1}{\sqrt{\cos\theta+s_2\sin\theta}}\right)\right].$$

**Table B1**. Calculated coefficients $b_{ij}$ and $B$ ($10^{-3}$/GPa) for different orientations. By defining the uniaxial strain rate $\dot{\varepsilon}_{yy} = 5.0 \times 10^8$ /s, the SIF rate can be evaluated as $\dot{K}_I = \sqrt{C'_{22}L_y/2B}\,\dot{\varepsilon}_{yy}$ in the unit of MPa$\sqrt{\text{m}}$/ps.

| orientations | $b_{11}$ | $b_{12}$ | $b_{22}$ | $b_{66}$ | $B$ | $C'_{22}$ (GPa) | $L_y$ (Å) | $\sqrt{C'_{22}L_y/2B}\,\dot{\varepsilon}_{yy}$ |
|---|---|---|---|---|---|---|---|---|
| (112)/[1$\bar{1}$0] | 3.3504 | -1.1356 | 3.7596 | 14.7246 | 4.3 | 310.2250 | 448.454 | 0.0201 |
| (111)/[1$\bar{1}$0] | 3.7596 | -1.1356 | 3.3504 | 14.7246 | 4.0 | 332.5133 | 466.211 | 0.0220 |
| (114)/[1$\bar{1}$0] | 3.9258 | -1.9837 | 4.8806 | 11.3319 | 4.4 | 270.6013 | 461.874 | 0.0188 |
| (221)/[1$\bar{1}$0] | 4.8806 | -1.9837 | 3.9258 | 11.3319 | 4.0 | 322.6074 | 446.589 | 0.0212 |
| (113)/[1$\bar{1}$0] | 3.6691 | -1.6403 | 4.4503 | 12.7058 | 4.4 | 285.3579 | 455.997 | 0.0192 |
| (332)/[1$\bar{1}$0] | 4.4503 | -1.6403 | 3.6691 | 12.7058 | 4.0 | 327.9083 | 456.574 | 0.0216 |
| (223)/[1$\bar{1}$0] | 3.3340 | -0.9507 | 3.4062 | 15.4642 | 4.2 | 324.8011 | 448.529 | 0.0208 |
| (334)/[1$\bar{1}$0] | 3.4062 | -0.9507 | 3.3340 | 15.4642 | 4.1 | 328.7343 | 467.503 | 0.0216 |
| (116)/[1$\bar{1}$0] | 4.1900 | -2.3197 | 5.2883 | 9.9880 | 4.4 | 256.8812 | 458.900 | 0.0183 |
| (552)/[1$\bar{1}$0] | 5.1334 | -2.1911 | 4.0876 | 10.5026 | 4.0 | 319.0303 | 462.838 | 0.0215 |
| (118)/[1$\bar{1}$0] | 4.3032 | -2.4604 | 5.4564 | 9.4252 | 4.4 | 251.2806 | 464.426 | 0.0182 |
| (335)/[1$\bar{1}$0] | 3.3097 | -0.9902 | 3.5096 | 15.3059 | 4.2 | 320.1698 | 450.710 | 0.0207 |

## Appendix C. correlation between $G_c$ and $K_{Ic}$

Since the stress and energy approaches in ***linear elastic fracture mechanics*** (***LEFM***) are equivalent, a unique relation between the energy release rate $G_c$ and stress intensity factor $K_{Ic}$ can be derived via the *crack closure method*. Considering a Mode-I crack before and after an extension distance d$a$, two corresponding coordinates *x*-*y* and *x*′-*y*′ centered at the crack-tip are established ($x' = x - \mathrm{d}a$, $y' = y$, $r^2 = x'^2 + y'^2$). According to the anisotropic ***LEFM*** stress fields, the normal stress ahead of the crack-tip ((*x*, *y*) = (*x*, 0) in current *x*-*y* coordinate) before extension can be obtained by inserting $r = x$ and $\theta = 0$ into Eq.(B.8),

$$\sigma_{yy} = \frac{K_I}{\sqrt{2\pi x}} \tag{C.1}$$

Assuming the crack propagation by a very small distance d$a$, the displacement along the *y*-axis direction of newly-formed crack surfaces ($0 \leq x \leq \mathrm{d}a$) can be obtained by inserting $r = |x'|$ into Eq.(B.7),

$$u_y = K_I\sqrt{\frac{2(\mathrm{d}a - x)}{\pi}}\,\Re\left[\frac{1}{s_1 - s_2}(s_1 q_2 - s_2 q_1)\right] \tag{C.2}$$

The strain energy associated with this process can be evaluated as the work done by $\sigma_{yy}$ before crack extension, which closes up the crack opening after the crack extension. The work done by transversing equals to the energy release rate under Mode-I extension,

$$G_c \mathrm{d}a = 2\int_0^{\mathrm{d}a} \frac{1}{2}\sigma_{yy} u_y \mathrm{d}x \tag{C.3}$$

By substituting Eq.(C.1) and Eq.(C.2) into Eq.(C.3), the formula can be simplified as,

$$G_c = \frac{K_{Ic}^2}{2}\Re\left[\frac{1}{s_1 - s_2}(s_1 q_2 - s_2 q_1)\right] \tag{C.4}$$

where and are obtained by solving the governing equation Eq.(B.3), especially which reduces to the following form when the material is of orthotropic symmetry with respect to *x*-*z* and *y*-*z* planes ($b_{16} = b_{26} = 0$),

$$b_{11}\mu_j^4 + (2b_{12} + b_{66})\mu_j^2 + b_{22} = 0 \tag{C.5}$$

Analytical solutions for $s_1$ and $s_2$ are thus obtained by solving Eq.(C.5),

$$\begin{aligned} s_1^2 = \mu_1^2 &= \frac{2b_{12} + b_{66}}{2b_{11}} + \frac{\sqrt{(2b_{12} + b_{66})^2 - 4b_{11}b_{22}}}{2b_{11}}, \\ s_2^2 = \mu_2^2 &= \frac{2b_{12} + b_{66}}{2b_{11}} - \frac{\sqrt{(2b_{12} + b_{66})^2 - 4b_{11}b_{22}}}{2b_{11}}. \end{aligned} \tag{C.6}$$

While $q_1$ and $q_2$ can be determined by Eq.(B.6), the energy release rate can be simplified as,

$$G_c = K_{Ic}^2 B \quad \text{(C.7)}$$

where,

$$B = \sqrt{\frac{b_{11}b_{22}}{2}\left(\frac{2b_{12}+b_{66}}{2b_{11}} + \sqrt{\frac{b_{22}}{b_{11}}}\right)} \quad \text{(C.8)}$$

The calculated $B$ values for 12 different orientations are listed in **Table B1**. According to the Griffith theory, the critical SIF $K_{IG}$ can be arrived by inserting the Eq.(3.4) and Eq.(3.5) into Eq.(C.7) for single crystal and bicrystal, respectively.

**Table C1**. Calculated cleavage SIF $K_{IG}$ for the intergranular fracture of bicrystalline specimens with 12 different orientations. The $K_{Ie}$ is further evaluated in **Table D1** in **Appendix D**.

| orientation | $\gamma_{GB}$ (J/m$^2$) | $\gamma_s^c$ (J/m$^2$) | $G_I^{gb}$ (J/m$^2$) | $B$ ($10^{-3}$/GPa) | $K_{IG}$ (MPa$\cdot\sqrt{\text{m}}$) | $K_{IG} - K_{Ie}$ |
|---|---|---|---|---|---|---|
| (112)/[1$\bar{1}$0] | 0.271 | 1.853 | 3.435 | 4.3 | 0.894 | > 0 |
| (111)/[1$\bar{1}$0] | 1.308 | 1.965 | 2.622 | 4.0 | 0.810 | < 0 |
| (114)/[1$\bar{1}$0] | 1.169 | 1.502 | 1.835 | 4.4 | 0.646 | > 0 |
| (221)/[1$\bar{1}$0] | 1.265 | 1.844 | 2.423 | 4.0 | 0.778 | > 0 |
| (113)/[1$\bar{1}$0] | 1.141 | 1.891 | 2.641 | 4.4 | 0.775 | > 0 |
| (332)/[1$\bar{1}$0] | 1.150 | 1.868 | 2.586 | 4.0 | 0.804 | > 0 |
| (223)/[1$\bar{1}$0] | 1.118 | 1.919 | 2.720 | 4.2 | 0.805 | > 0 |
| (334)/[1$\bar{1}$0] | 1.208 | 1.866 | 2.524 | 4.1 | 0.785 | > 0 |
| (116)/[1$\bar{1}$0] | 1.154 | 1.872 | 2.590 | 4.4 | 0.767 | > 0 |
| (552)/[1$\bar{1}$0] | 1.290 | 1.852 | 2.414 | 4.0 | 0.777 | > 0 |
| (118)/[1$\bar{1}$0] | 1.106 | 1.836 | 2.566 | 4.4 | 0.764 | > 0 |
| (335)/[1$\bar{1}$0] | 0.952 | 1.908 | 2.864 | 4.2 | 0.826 | > 0 |

## Appendix D. Rice criterion for dislocation emission in anisotropic materials

The classical ***Rice theory*** based on the ***Peierls concept*** to predict the dislocation nucleation from a crack-tip in isotropic materials [12] can be generalized to full anisotropic elasticity as [14, 88],

$$G = \boldsymbol{K}^T \boldsymbol{\Lambda} \boldsymbol{K} \quad \text{(D.1)}$$

where, $\boldsymbol{K} = [K_{II} \quad K_I \quad K_{III}]^T = [K_x \quad K_y \quad K_z]^T$ is the SIF matrix for a mixed loading mode, and $\boldsymbol{\Lambda}$ is the Stroh energy tensor,

$$\boldsymbol{\Lambda} = \frac{1}{2}\Re(i\boldsymbol{A}\boldsymbol{B}^{-1}) \tag{D.2}$$

where, $\boldsymbol{A} = [\boldsymbol{a}_1 \quad \boldsymbol{a}_2 \quad \boldsymbol{a}_3]$ and $\boldsymbol{B} = [\boldsymbol{b}_1 \quad \boldsymbol{b}_2 \quad \boldsymbol{b}_3]$ are complex matrices composed of eigenvectors $\boldsymbol{a}$ and $\boldsymbol{b}$, and satisfy the following eigenvalue equation,

$$\boldsymbol{N}\begin{bmatrix} \boldsymbol{A} & \overline{\boldsymbol{A}} \\ \boldsymbol{B} & \overline{\boldsymbol{B}} \end{bmatrix} = \begin{bmatrix} \boldsymbol{A} & \overline{\boldsymbol{A}} \\ \boldsymbol{B} & \overline{\boldsymbol{B}} \end{bmatrix}\begin{bmatrix} \boldsymbol{P} & \boldsymbol{0} \\ \boldsymbol{0} & \overline{\boldsymbol{P}} \end{bmatrix} \tag{D.3}$$

where, $\boldsymbol{P} = \langle p_* \rangle = \mathrm{diag}[p_1, p_2, p_3]$ are eigenvalues defined in the Stroh formalism, and $\boldsymbol{N}$ is the ***fundamental elasticity matrix*** [89] defined as,

$$\boldsymbol{N} = \begin{bmatrix} \boldsymbol{N}_1 & \boldsymbol{N}_2 \\ \boldsymbol{N}_3 & \boldsymbol{N}_1^T \end{bmatrix} \tag{D.4}$$

with

$$\begin{cases} \boldsymbol{N}_1 = -\boldsymbol{T}^{-1}\boldsymbol{R}^T \\ \boldsymbol{N}_2 = \boldsymbol{T}^{-1} \\ \boldsymbol{N}_3 = \boldsymbol{R}\boldsymbol{T}^{-1}\boldsymbol{R}^T - \boldsymbol{Q} \end{cases} \tag{D.5}$$

where,

$$\boldsymbol{Q} = \begin{bmatrix} C'_{11} & C'_{16} & C'_{15} \\ C'_{16} & C'_{66} & C'_{56} \\ C'_{15} & C'_{56} & C'_{55} \end{bmatrix}, \boldsymbol{R} = \begin{bmatrix} C'_{16} & C'_{12} & C'_{14} \\ C'_{66} & C'_{26} & C'_{46} \\ C'_{56} & C'_{25} & C'_{45} \end{bmatrix}, \boldsymbol{T} = \begin{bmatrix} C'_{66} & C'_{26} & C'_{46} \\ C'_{26} & C'_{22} & C'_{24} \\ C'_{46} & C'_{24} & C'_{44} \end{bmatrix} \tag{D.6}$$

with the stiffness tensor components $C'_{ij}$ given by Eq.(A.5).

Considering the emission of a dislocation with the Burgers vector at an angle $\phi$ with respect to a vector lying on the slip plane and perpendicular to the crack-front direction, and occurring along a slip plane at an angle $\theta$ to the crack plane, the critical Mode-I SIF is,

$$K_{Ie} = \frac{\sqrt{G_{Ie}o(\theta,\phi)}}{F_{xy}(\theta)\cos\phi} \tag{D.7}$$

where, $G_{Ie}$ is the critical energy release rate for dislocation emission [20],

$$G_{Ie} = \begin{cases} 0.145\gamma_s^e + 0.5\gamma_{usf} & \gamma_s^e > 3.45\gamma_{usf} \\ \gamma_{usf} & \gamma_s^e < 3.45\gamma_{usf} \end{cases} \tag{D.8}$$

with $\gamma_s^e$ as the surface energy for the emission plane, and $F_{xy}(\theta)$ is a geometrical factor,

$$\frac{K_I}{\sqrt{2\pi r}}F_{xy}(\theta) = \left(\sigma_{yy} - \sigma_{xx}\right)\sin\theta\cos\theta + \sigma_{xy}(\cos^2\theta - \sin^2\theta)$$
$$\therefore F_{xy}(\theta) = \begin{matrix} \sin\theta\cos\theta\Re\left[\frac{1}{s_1-s_2}\left(\frac{s_1(1-s_2^2)}{\sqrt{\cos\theta+s_2\sin\theta}} - \frac{s_2(1-s_1^2)}{\sqrt{\cos\theta+s_1\sin\theta}}\right)\right] \\ +(\sin^2\theta - \cos^2\theta)\Re\left[\frac{s_1 s_2}{s_1-s_2}\left(\frac{1}{\sqrt{\cos\theta+s_2\sin\theta}} - \frac{1}{\sqrt{\cos\theta+s_1\sin\theta}}\right)\right] \end{matrix} \quad \text{(D.9)}$$

with $\sigma_{ij}$ $(i, j = x, y)$ given by Eq.(B.8), and the elasticity coefficient $o(\theta, \phi)$ is,

$$o(\theta, \phi) = s_i(\phi)\left[\Lambda_{ij}^{\theta}\right]^{-1} s_j(\phi) \quad \text{(D.10)}$$

with

$$\boldsymbol{s}(\phi) = [\cos\phi \quad 0 \quad \sin\phi] \quad \text{(D.11)}$$

and

$$\Lambda_{ij}^{\theta} = \Omega_{ik}\Lambda_{kl}\Omega_{jl} \quad \text{(D.12)}$$

where the rotation matrix $\boldsymbol{\Omega}$ is,

$$\boldsymbol{\Omega} = \begin{bmatrix} \cos\theta & \sin\theta & 0 \\ -\sin\theta & \cos\theta & 0 \\ 0 & 0 & 1 \end{bmatrix} \quad \text{(D.13)}$$

As suggested by Weinberger *et al*. [90], the fundamental slip planes of $\alpha$-iron are $\{110\}$ and $\{112\}$ planes. We thus calculate, in **Table D1**, the maximum Schmid factor from all 24 slip systems [91] in the grain #1 of the bicrystalline specimen shown in **Fig. 1**. The calculated unstable stacking fault energies $\gamma_{usf}$ and surface energies $\gamma_s^e$ for slip along $\langle 111 \rangle$ directions in $\{110\}$ and $\{112\}$ slip planes are in fairly good agreement with previous studies [75], where various many-body potentials are evaluated.

**Table D1**. Calculated critical SIF $K_{Ie}$ (MPa·$\sqrt{\text{m}}$) in the single crystal (*i.e.* the grain #1 or grain #2 in **Fig. 1**a in the main text) for 12 different orientations. The elastic coefficient $o(\theta,\phi)$ is in the unit of GPa. All energy quantities, including the unstable stacking fault energy $\gamma_{usf}$, the surface energy for the emission plane $\gamma_s^e$, the surface energy for the single grain #1 in the bicrystalline specimens $\gamma_s^c$, and the critical energy release rate for dislocation emission $G_{Ie}$, are in the unit of J/m$^2$.

| orientation | Sch. fac. | slip $(hkl)[uvw]$ | $\theta$ | $\phi$ | $F_{xy}(\theta)$ | $o(\theta,\phi)$ | $\gamma_{usf}$ | $\gamma_s^e$ | $G_{Ie}$ | $K_{Ie}$ |
|---|---|---|---|---|---|---|---|---|---|---|
| $(112)/[1\bar{1}0]$ | 0.465 | $(110)[\bar{1}11]$ | 54.7° | 54.7° | 0.8825 | 185.8419 | 0.624 | 1.621 | 0.624 | 0.668 |
| $(111)/[1\bar{1}0]$ | 0.465 | $(110)[\bar{1}11]$ | 35.3° | 54.7° | 0.7232 | 185.8419 | 0.624 | 1.621 | 0.624 | 0.815 |
| $(114)/[1\bar{1}0]$ | 0.479 | $(10\bar{1})[111]$ | 60.0° | 0.0° | 0.9033 | 246.1671 | 0.624 | 1.621 | 0.624 | 0.434 |
| $(221)/[1\bar{1}0]$ | 0.495 | $(\bar{2}11)[111]$ | 65.9° | 0.0° | 0.8313 | 244.6089 | 0.767 | 1.853 | 0.767 | 0.521 |
| $(113)/[1\bar{1}0]$ | 0.461 | $(10\bar{1})[111]$ | 64.8° | 0.0° | 0.8726 | 246.6791 | 0.624 | 1.621 | 0.624 | 0.450 |
| $(332)/[1\bar{1}0]$ | 0.496 | $(\bar{2}11)[111]$ | 64.2° | 0.0° | 0.8377 | 240.7773 | 0.767 | 1.853 | 0.767 | 0.513 |
| $(223)/[1\bar{1}0]$ | 0.492 | $(110)[\bar{1}11]$ | 46.7° | 54.7° | 0.8430 | 186.9028 | 0.624 | 1.621 | 0.624 | 0.701 |
| $(334)/[1\bar{1}0]$ | 0.492 | $(110)[\bar{1}11]$ | 43.3° | 54.7° | 0.8149 | 186.9028 | 0.624 | 1.621 | 0.624 | 0.725 |
| $(116)/[1\bar{1}0]$ | 0.487 | $(11\bar{2})[111]$ | 48.5° | 0.0° | 0.8963 | 248.7967 | 0.763 | 1.853 | 0.763 | 0.486 |
| $(552)/[1\bar{1}0]$ | 0.485 | $(\bar{1}21)[11\bar{1}]$ | 67.1° | 0.0° | 0.8257 | 258.2007 | 0.770 | 1.853 | 0.770 | 0.540 |
| $(118)/[1\bar{1}0]$ | 0.490 | $(11\bar{2})[111]$ | 45.3° | 0.0° | 0.8783 | 248.7967 | 0.763 | 1.853 | 0.763 | 0.496 |
| $(335)/[1\bar{1}0]$ | 0.486 | $(110)[\bar{1}11]$ | 49.7° | 54.7° | 0.8623 | 186.6757 | 0.624 | 1.621 | 0.624 | 0.685 |

## Appendix E. procedure for fitting MD results to obtain SIFs

As shown in following **Fig. E1**, the rectangular region containing the GB is divided into a series of bins, in which the atomic stress components are recorded and outputted during simulations. However, since the central region containing the crack is also outputted, the corresponding data should be excluded. Specifically, the post-process code initially locates the center of the original crack along the x-axis direction, then moves towards both the right- and left-directions to find the nearest positions with local maximum stress, i.e. the crack-tips. Once the two crack-tip positions were determined, the data points within this regime (indicated by the green dash rectangular in **Fig. E1**) would be deleted. Finally, the remaining two parts of data points were fitted to the fracture mechanics formula, respectively.

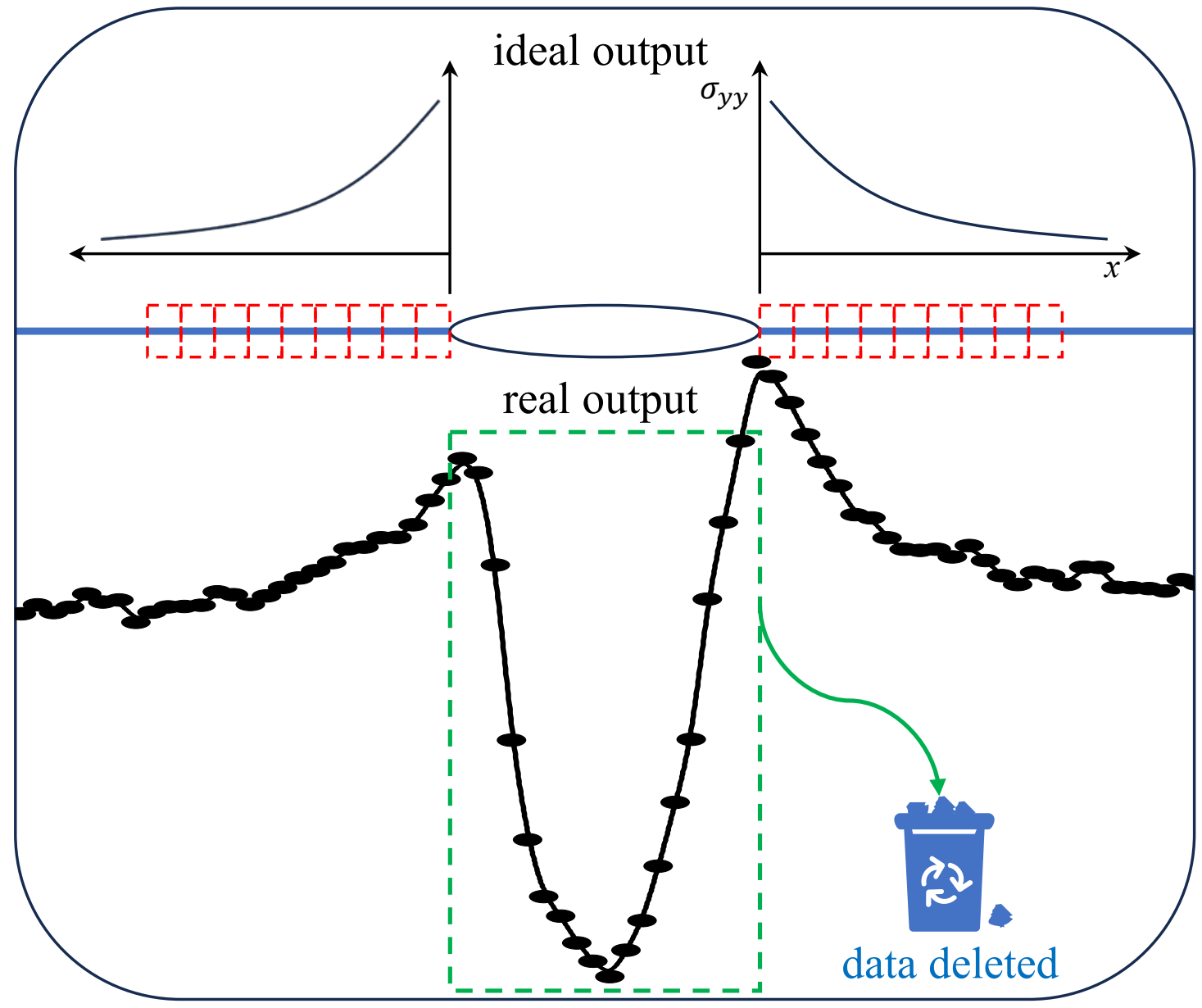

**Fig. E1**. Schematics of fitting MD results to obtain SIFs.

## Appendix F. atomic illustration of the GB transition

Without loss of generality, in **Fig. F1** we show the atomic evolution of GB structural unit of Σ9(38.90) and Σ11(50.50) GBs. It is found that the pearl structure of Σ9(38.90) GB and the kite structure of Σ11(50.50) GB both evolves into the coral-like structure under uniaxial tension. As shown in **Fig. 6** & **7** in the main text, this transition is induced by the twin growth aside the GB.

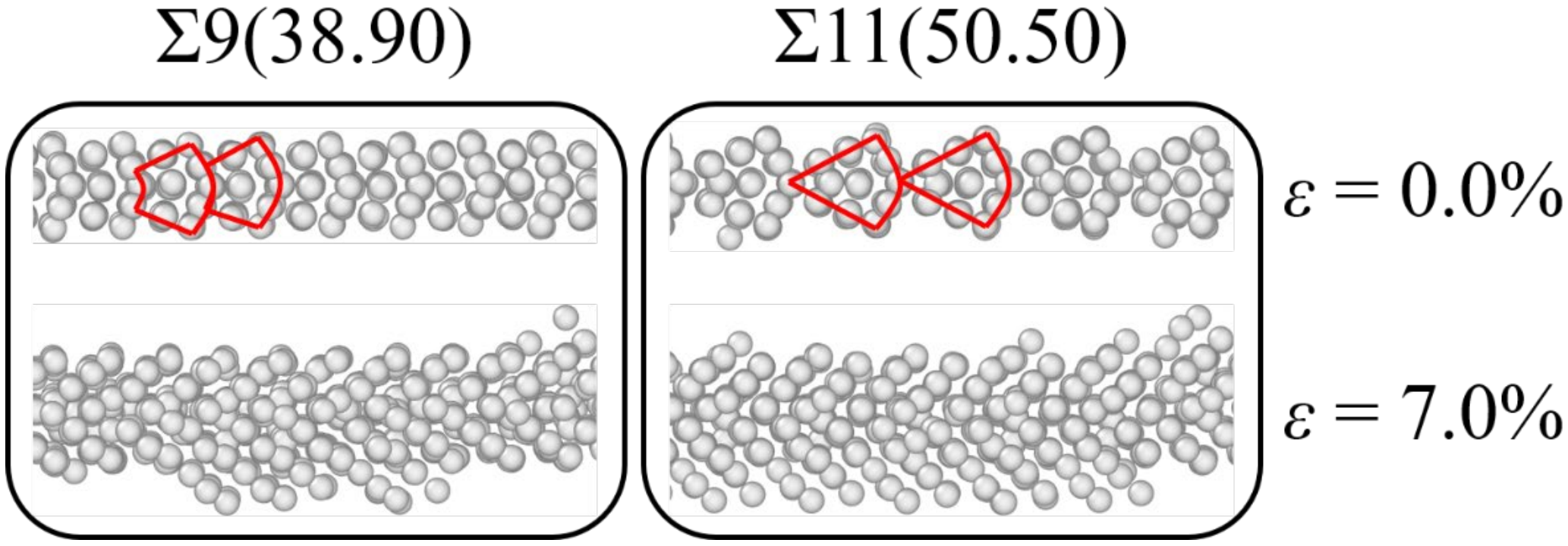

**Fig. F1**. Atomic details of the GB structural transition.